\documentclass[main=english,italian]{aa}  

\usepackage{graphicx}
\usepackage{booktabs,caption}
\usepackage{makecell}
\usepackage{txfonts}
\begin{document} 

   \title{The ASTRODEEP-GS43 catalogue: new photometry and redshifts for the CANDELS GOODS-South field}

   \author{E. Merlin\inst{1} \email{emiliano.merlin@inaf.it}
          \and
          M. Castellano\inst{1}
          \and
          P. Santini\inst{1}
          \and
          G. Cipolletta\inst{1}
          \and
          K. Boutsia\inst{2}
          \and
          C. Schreiber\inst{3}
          \and
          F. Buitrago\inst{4}
          \and
          A. Fontana\inst{1}
          \and
          D. Elbaz\inst{5}
          \and
          J. Dunlop\inst{6}
          \and
          A. Grazian\inst{7}
          \and 
          R. McLure\inst{6}
          \and
          D. McLeod\inst{6}
          \and
          M. Nonino\inst{8}
          \and
          B. Milvang-Jensen\inst{9}          
          \and
          S. Derriere\inst{10}  
          \and
          N. P. Hathi\inst{11}
          \and
          L. Pentericci\inst{1}
          \and
          F. Fortuni\inst{1}
          \and
          A. Calabrò\inst{1}
          }

   \institute{
              $^{1}$INAF - OAR, via Frascati 33, 00078 Monte Porzio Catone (Roma) - Italy\\
              $^{2}$Las Campanas Observatory, Carnegie Observatories, Colina El Pino, Casilla 601, La Serena, Chile\\
              $^{3}$Department of Physics, University of Oxford, Keble Road, Oxford, OX1 3RH, UK\\
              $^{4}$Instituto de Astrofísica e Ciências do Espaço, Universidade de Lisboa, Portugal\\
              $^{5}$Laboratoire AIM-Paris-Saclay, CEA/DRF/Irfu - CNRS - Université Paris  Diderot, CEA-Saclay, pt courrier 131, 91191, Gif-sur-Yvette, France\\
              $^{6}$Institute for Astronomy, University of Edinburgh, Royal Observatory, Edinburgh EH9 3HJ, UK\\
              $^{7}$INAF-Osservatorio Astronomico di Padova, Vicolo dell'Osservatorio 5, I-35122, Padova, Italy\\
              $^{8}$INAF - Osservatorio Astronomico di Trieste, Via Tiepolo 11, I-34131 Trieste, Italy\\
              $^{9}$Cosmic Dawn Center (DAWN); Niels Bohr Institute, University of Copenhagen, Lyngbyvej 2, DK-2100 Copenhagen, Denmark\\
              $^{10}$Université de Strasbourg, CNRS, Observatoire astronomique de Strasbourg, Strasbourg, 67000, France\\
              $^{10}$Space Telescope Science Institute, 3700 San Martin Drive, Baltimore, MD 21218, USA
             }

   \date{08/01/2021}
    \titlerunning{The ASTRODEEP-GS43 Catalogue}
    \authorrunning{E. Merlin et al.}

 
  \abstract
   {We present ASTRODEEP-GS43, a new multiwavelength photometric catalogue of the GOODS-South field, which builds and improves upon the previously released CANDELS catalogue.}
   {We provide photometric fluxes and corresponding uncertainties in 43 optical and infrared bands (25 wide and 18 medium filters), as well as photometric redshifts and physical properties of the 34930 CANDELS $H$-detected objects, plus an additional sample of 178 $H$-dropout sources, of which 173 are $Ks$-detected and 5 IRAC-detected.}
   {We keep the CANDELS photometry in 7 bands (CTIO $U$, \textit{Hubble} Space Telescope WFC3 and ISAAC-$K$), and measure from scratch the fluxes in the other 36 (VIMOS, $HST$ ACS, HAWK-I $Ks$, \textit{Spitzer} IRAC, and 23 from \textit{Subaru} SuprimeCAM and \textit{Magellan Baade} Fourstar) with state-of-the-art techniques of template-fitting. We then compute new photometric redshifts with three different software tools, and take the median value as best estimate. We finally evaluate new physical parameters from SED fitting, comparing them to previously published ones.}
   {Comparing to a sample of 3931 high quality spectroscopic redshifts, for the new photo-$z$'s we obtain a normalized median absolute deviation (NMAD) of 0.015, with 3.01\% of outliers on the full catalogue (0.011, 0.22\% on the bright end at $I814$<22.5), similarly to the best available published samples of photometric redshifts, such as the COSMOS UltraVISTA catalogue.}
   {The ASTRODEEP-GS43 results are in qualitative agreement with previously published catalogues of the GOODS-South field, improving on them particularly in terms of SED sampling and photometric redshift estimates. The catalogue is available for download from the \textsc{Astrodeep} website.}


   \maketitle
%

\section{Introduction}

\begin{figure*}
\centering
\includegraphics[width=18cm]{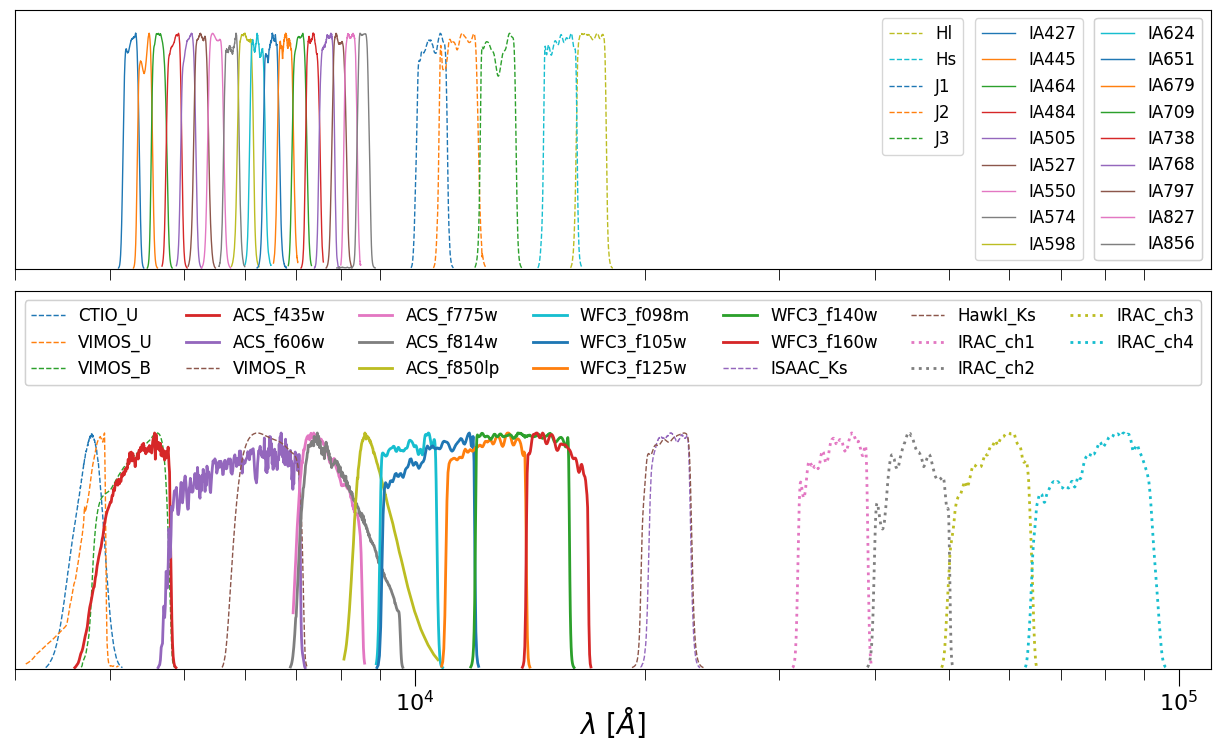}
\caption{The complete passbands set for the ASTRODEEP-GS43 catalogue. The upper panel shows the filters of the \textsc{MUSYC} and \textsc{ZFourge} medium bands, while the lower panel shows the wide bands from $HST$, \textit{Spitzer}, and ground based facilities; the curves are normalized to arbitrary units to make them peak at unitary transmission.} \label{filters}
\end{figure*}

\begin{table}
\renewcommand{\arraystretch}{1.5}
\caption{Summary of the 20 wide bands in the catalogue.}
\centering
\begin{tabular}{ | l | l | l | l | l | l |}
\hline
Instrument & Filter & \makecell{$\lambda_{ref}$ \\ (nm)} & \makecell{$\Delta \lambda$ \\ (nm)}  & \makecell{PSF \\ ('')} & \makecell{5$\sigma$ \\ Depth \\ (AB)} \\ \hline\hline
\makecell{$Blanco$ \\ MOSAIC \\ II} & CTIO $U$ & 358.4 & 62.5 & 1.37 & 26.63$^a$ \\ \hline
\makecell{$VLT$ \\ VIMOS} & $U$ & 371.2 & 38.0 & 0.80 & 28.21$^b$ \\ 
 & $B$ &  427.6 & 96.3 & 0.85 & 28.74$^b$ \\ 
 & $R$ &  641.4 & 135.0 & 0.75 & 27.96$^b$ \\ \hline
\makecell{$HST$ \\ ACS} & $F435W$ & 432.9 & 93.9 & 0.08 & 28.83$^c$ \\ 
 & $F606W$ & 592.2 & 232.3 & 0.08 & 29.24$^c$ \\ 
 & $F775W$ & 769.3 & 151.1 & 0.08 & 28.48$^c$ \\ 
 & $F814W$ & 811.6 & 230.3 & 0.09 & 29.35$^c$ \\ 
 & $F850LP$ & 914.4 & 148.9 & 0.09 & 28.54$^c$ \\ \hline
\makecell{$HST$ \\ WFC3} & $F098M$ & 986.3 & 169.4 & 0.13 & 28.18$^c$ \\ 
 & $F105W$ & 1055.0 & 291.7 & 0.15 & 28.70$^c$ \\ 
 & $F125W$ & 1248.6 & 300.5 & 0.16 & 28.85$^c$ \\ 
 & $F140W$ & 1392.3 & 394.1 & 0.17 & 27.64$^c$ \\ 
 & $F160W$ & 1537.0 & 287.4 & 0.17 & 28.72$^c$  \\ \hline
\makecell{VLT \\ ISAAC} & $Ks$ & 2159.2 & 274.6 & 0.48 & 25.09$^d$ \\ \hline
\makecell{VLT \\ HAWK-I} & $Ks$ & 2142.0 & 325.0 & 0.43 & 26.26$^b$ \\ \hline
\makecell{\textit{Spitzer} \\ IRAC} & CH1 & 3537.8 & 743.2 & 1.66 & 25.63$^b$ \\ 
& CH2 & 4478.0 & 1009.7 & 1.72 & 25.51$^b$ \\ 
& CH3 & 5696.2 & 1391.2 & 1.88 & 23.28$^b$ \\ 
& CH4 & 7797.8 & 2831.2 & 1.98 & 23.16$^b$ \\ \hline
\end{tabular} \label{filterstab1}
\small \\ $^a$ Aperture magnitude within a radius 1 FWHM of the PSF.\\ $^b$ Median total magnitude at 5$\sigma$; the given values are averages of the varying depths in the field. \\ $^c$ Median aperture magnitudes within a fixed radius of 0.17''; the given values are averages of the varying depths in the field (including CANDELS-deep field and the HUDF depths). \\ $^d$ PSF and depth vary among ISAAC tiles, the value is the median of all available tiles. 
\end{table} 

\begin{table}
\renewcommand{\arraystretch}{1.5}
\caption{Summary of the 23 ground-based medium bands in the catalogue.}
\centering
\begin{tabular}{ | l | l | l | l | l | l |}
\hline
Instrument & Filter & \makecell{$\lambda_{ref}$ \\ (nm)} & \makecell{$\Delta \lambda$ \\ (nm)} & \makecell{PSF  \\ ('')} & \makecell{5$\sigma$ \\ Depth \\ (AB)$^a$} \\ \hline\hline
\makecell{$Subaru$ \\ SuprimeCAM} & $IA427$ & 427.0 & 20.7 & 1.01 & 25.57 \\ 
& $IA445$ & 445.0 & 20.0 & 1.23 & 25.91 \\ 
& $IA464$ & 464.0 & 22.0 & 1.79 & 25.05 \\ 
& $IA484$ & 484.0 & 23.0 & 0.76 & 26.69 \\ 
& $IA505$ & 505.0 & 26.0 & 0.94 & 25.85 \\ 
& $IA527$ & 527.0 & 24.0 & 0.83 & 26.63 \\ 
& $IA550$ & 550.0 & 28.0 & 1.13 & 26.04 \\ 
& $IA574$ & 574.0 & 27.0 & 0.95 & 25.56 \\ 
& $IA598$ & 598.0 & 30.0 & 0.63 & 26.78 \\ 
& $IA624$ & 624.0 & 30.0 & 0.61 & 26.59 \\ 
& $IA651$ & 651.0 & 33.0 & 0.60 & 26.91 \\ 
& $IA679$ & 679.0 & 34.0 & 0.80 & 26.59 \\ 
& $IA709$ & 709.0 & 32.0 & 1.60 & 25.32 \\ 
& $IA738$ & 738.0 & 33.0 & 0.77 & 26.54 \\ 
& $IA767$ & 767.0 & 37.0 & 0.70 & 25.27 \\ 
& $IA797$ & 797.0 & 35.0 & 0.68 & 25.25 \\ 
& $IA827$ & 827.0 & 34.0 & 1.69 & 24.58 \\ 
& $IA856$ & 856.0 & 34.0 & 0.67 & 24.89 \\ \hline
\makecell{$Magellan$\\$Baade$\\ FourStar} & $J1$ & 1054.0 & 103.0 & 0.59 & 25.86 \\ 
& $J2$ & 1144.8 & 141.0 & 0.62 & 25.71 \\ 
& $J3$ & 1280.2 & 132.0 & 0.56 & 25.70 \\ 
& $Hs$ & 1554.4 & 160.0 & 0.60 & 24.99 \\ 
& $Hl$ & 1702.0 & 161.0 & 0.50 & 25.28 \\\hline
\end{tabular} \label{filterstab2}
\small \\ $^a$ Median total magnitude at 5$\sigma$; the given values are averages on the varying depth in the field.
\end{table}

Multi-wavelength extragalactic astronomy targeting the high-redshift Universe has matured to the status of precision science during the past decade. 
Deep optical/infrared photometric surveys like CANDELS \citep{Grogin2011,Koekemoer2011}, 3D-HST \citep{Skelton2014} or Frontier Fields \citep{Lotz2014,Koekemoer2014,Merlin2016b,Castellano2016,Marchesini2018} have provided high quality multi-band imaging of various regions of the sky, combining data from space observatories (the \textit{Hubble} and \textit{Spitzer} Space Telescopes) with data from ground-based facilities (ESO Very Large Telescope, Keck, Subaru), extending the spectroscopic limit and allowing for the analysis of statistically significant samples of galaxies up to $z \sim 8-9$. These projects paved the way for the upcoming generation of observational campaigns and instruments, including large-scale surveys and deep imaging programs (with Vera Rubin Observatory / LSST, Euclid, James Webb and Nancy Grace Roman Space Telescopes), or the extremely high resolution power provided by Adaptive Optics (e.g. the Extremely Large Telescope MICADO instrument). The advent of such new facilities will further push the limits of our observations towards the earliest epochs of structure formation. Waiting for such exciting game changing technologies, there is still room to exploit the currently available data combining all the archival observations to extract as much information as possible.

The CANDELS legacy stands among the most informative collections of extragalactic data. In particular, the Great Observatories Origins Deep Survey Field South (GOODS-South, GS hereafter) represents a benchmark for its exquisite quality and its richness. Located at RA = 3h 32m 30s, dec = -27$^{\circ}$ 48m 20s, with an area of $\sim$ 173 sq. arcmin, GS has been observed by a number of observatories both from the ground and from space, at all wavelengths from X-rays to radio \citep[e.g. ][]{Elbaz2011,Ashby2013,Luo2017,Franco2018}, and it is worth mentioning here that the field will be the target of the James Webb GTO program \textit{JADES} (P.I. Rieke and Ferruit). The first generation official CANDELS catalogue by \citet[][G13 hereafter]{Guo2013} includes 34930 galaxies, with photometric data in 17 wide pass-bands, from the ultraviolet (UV) to the mid-infrared, adding then-new observations from $HST$ WFC3 to existing archival images: 9 $HST$ bands (from the ACS and WFC3 cameras) and 4 IRAC bands from \textit{Spitzer}, plus three $VLT$ bands (VIMOS U, ISAAC $K$ and HAWK-I $Ks$) and an additional $U$ band from the CTIO MOSAIC instrument. The detection was performed on the WFC3 $H$160 band, using \textsc{SExtractor} \citep{Bertin1996} in a dual ``hot+cold'' mode \citep{Galametz2013}; photometric measurements were performed using again \textsc{SExtractor} on the $HST$ images, to measure a total magnitude on the detection band, and PSF-matched isophotal colors for the other bands; while a template-fitting technique was used to directly estimate total fluxes on ground-based and IRAC images, with the code \textsc{T-FIT} \citep{Laidler2007}. The 3D-HST catalogue by \citet{Skelton2014} also used part of this first collection of data.
 
Subsequently, two additional bands (VIMOS $B$ and WFC3 $J$140) and deeper $Ks$ photometry from HAWK-I \citep[HUGS survey, ][]{Fontana2014,Grazian2015} have been added to the archival data, and new $HST$-ACS deep mosaics were released by the \textit{Hubble Legacy Fields} project\footnote{https://archive.stsci.edu/prepds/hlf/} \citep[HLF,][]{Illingworth2016, Whitaker2019}. 
Furthermore, deeper mosaics on 3.6 and 4.5 $\mu$m IRAC channels were created by R. McLure and used by \citet{Merlin2018,Merlin2019}, although no catalogue was released for the latter. 

This new generation of data gathered since 2013, along with the introduction of new techniques and software such as \textsc{t-phot} \citep{Merlin2015,Merlin2016a}, sparked the necessity of reviewing the original catalogue. 
In this paper we present ASTRODEEP-GS43, a new photometric and photo-$z$ catalogue for GS intended to yield a comprehensive set of optical/NIR photometric information on the field before the advent of the upcoming next generation datasets based on new instruments. This release builds on the previously published CANDELS catalogue, and complements similar recently published efforts such as the 3D-HST catalogue by \citet{Skelton2014} and the HLF catalogue by \citet{Whitaker2019}. 

We summarize here the main improvements with respect to the previous CANDELS catalogue. While we keep the G13 detection list on the WFC3 $H$160 band, 
\begin{itemize}
    \item we added photometric measurements on 18 \textit{Subaru} SuprimeCAM medium bands \citep[][MUSYC catalogue]{Cardamone2010}, 5 medium bands from \textit{Magellan Baade} FourStar \citep[][ZFOURGE survey]{Straatman2016}, and the $B$ and $R$ bands from VIMOS, to the previously released 18 wide bands; so that the total number of pass-bands is now 43. They are described in Sect. \ref{dataset}.
    \item we added to the catalogue 173 new objects detected in the HUGS HAWK-I $Ks$ image, as described in Sect. \ref{kdet}. Furthermore, we added 5 more sources detected in the \textit{Spitzer} IRAC 3.6 and 4.5 $\mu$m (CH1 and CH2) bands. Of the latter, 3 are from the list of 10 $H$-dropouts by \citet[][W16 hereafter]{Wang2016}, the remaining 7 of their list being included in our $Ks$-detections list. The final additional 2 sources were again found by Wang and collaborators (priv. comm.), but they were excluded from their published list (see Sect. \ref{iracdet});
    \item we measured $HST$ ACS fluxes from the new, deep mosaics released by the HLF project, again using \textsc{SExtractor} to extract isophotal aperture PSF-matched photometry. We point out that the latest HLF data release (v2.0) includes photometric data on UV bands; however, when we compiled our catalogue the latest available release was v1.5, which did not include UV images. For this reason, ASTRODEEP-GS43 does not include UV data. We do not consider this a major drawback, since the main focus of the present release is on high-redshift galaxies, for which UV observations are not crucial;
    \item we exploited the template-fitting code \textsc{t-phot} v2.0 to extract new photometry from the images of ground-based medium bands and \textit{Spitzer} bands, using three substantial algorithmic improvements with respect to the standard methods of v1.0: (i) the use of priors from the closest-wavelength high-resolution band for all the medium bands, as opposed to just use priors from the $H$ detection band; (ii) the background subtraction during the fitting process and (iii) individual locally variable PSFs for IRAC \citep[these techniques are presented in][and are summarized in Sect. \ref{photometry}]{Merlin2016a}. Also, two of the IRAC images (CH1 and CH2) are the new mosaics, which are deeper than the ones used for the CANDELS release, so that we gain a substantial amount of detections previously catalogued as upper limits.
\end{itemize}

The paper is structured as follows. In Section \ref{dataset} we describe the full dataset. In Section \ref{detection} we focus on the detection techniques adopted to single out $H$ band dropouts on the $Ks$-band image. In Section \ref{photometry} we discuss the photometric methods adopted on the new images, particularly on the 20 medium bands. In Section \ref{redshifts} we present the new estimated photometric redshifts and physical properties, also comparing our results to previous ones and showing some diagnostic plots. Finally, in Section \ref{summary} we summarize the work and discuss some conclusions and possible future developments. Throughout the paper we adopt AB magnitudes \citep{Oke1983} and standard cosmological parameters ($H_0$=70.0 km/s/Mpc, $\Omega_{\Lambda}$=0.7, $\Omega_{m}$=0.3).

\section{Dataset} \label{dataset}

The catalogue includes photometric data on 43 bands, which we describe in this Section. The full list of the filters is given in Fig. \ref{filters} and Tables \ref{filterstab1} and \ref{filterstab2}, with data taken from the SVO Filter Service Profile website  \citep{Rodrigo2012}\footnote{http://svo2.cab.inta-csic.es/theory/fps/}.

\subsection{HST bands}

We include 10 bands from $HST$: five from ACS (F435W, F606W, F775W, F814W, F850LP) and five from WFC3 (F098M, F105W, F125W, F140W, F160W). As already mentioned, the ACS images are the new mosaics released by the HLF project (v1.5) and obtained adding to the archival CANDELS images all publicly available $HST$ observations on the $Chandra$ Deep Field South region. The improvement with respect to CANDELS ACS mosaics is particularly significant in the $I$814 band, where the exposure time per pixel is increased by a factor of $\gtrsim$2-3, depending on position. The WFC3 images are the same used for the official CANDELS survey catalogue of GOODS-South by \citet{Guo2013}. Full details on data processing and mosaic preparation are provided in \citet{Illingworth2016, Whitaker2019} and in \citet{Koekemoer2011} for ACS and WFC3 bands, respectively.

\subsection{VLT}

The catalogue includes the three \textit{VLT} VIMOS bands $U$, $B$ and $R$ \citep{Nonino2009}. The $U$ band was already included in the original CANDELS catalogue, however we re-measured the fluxes on it using \textsc{t-phot}. On the contrary, the $B$ and $R$ bands data had never been included in previous catalogues.

\subsection{\textit{Subaru} SuprimeCAM and \textit{Magellan} FourStar bands}

One of the most important additions to this release is the photometric measurements for 18 medium bands (typical width 20-30 nm) from the \textit{Subaru} SuprimeCAM \citep{Cardamone2010} dataset. We also included new measurements on the 5 \textit{Magellan Baade} FourStar bands \citep{Straatman2016}. We stress that we did not directly include in our catalogue any previously published photometric measurement; rather, we used \textsc{t-phot} to obtain consistent and homogeneous photometry across the whole spectrum (see Section \ref{photometry}).

\subsection{Spitzer}

We used IRAC CH1 and CH2 deep mosaics by R. McLure (priv. comm.), obtained combining images from seven observational programs \citep[Dickinson, van Dokkum, Labb\'e, Bouwens, and three by Fazio including SEDS and S-CANDELS: see][for details]{Ashby2015} into two \textit{supermaps}; they are equivalent to the ones by \citet{Labbe2015}, and reach an average depth of $\sim25.6$ (total magnitude at 5$\sigma$) on both channels. On the other hand, CH3 and CH4 are the same mosaics used for G13. We used \textsc{t-phot} on all the IRAC bands; the photometric measurements on these mosaics have already been used in our recent works on passive galaxies in the early Universe \citep{Merlin2018,Merlin2019}.

\section{Detection of $H$-dropouts} \label{detection}

We kept the G13 $H$ band list of 34930 sources as a baseline for our catalog; the reader can find the details about the detection process in the original paper. On top of that, we added 173 sources detected on the deep HAWK-I $Ks$ band (IDs from 34931 to 35103); we describe the detection method in the following subsection. Finally, we added 5 IRAC-detected sources (IDs from 35104 on), from the W16 study (see Sect. \ref{iracdet} for details). In total, we therefore have 35108 catalogued objects. 

\subsection{K-selected sources} \label{kdet}

We used two different approaches to detect sources on the $Ks$ image, while excluding entries already present in G13.

\begin{itemize}
\item \textbf{Method 1 - Source Extractor in dual mode and cross-matching with G13}.
This method is aimed at detecting isolated sources that have escaped detection in the $H$ band, typically because they are too faint (below the 5$\sigma$ level) but visible in $K$. We resampled the HUGS HAWK-I image (original pixel scale 0.1065'') to the $HST$ mosaics pixel scale (0.06'') using \textsc{Swarp} \citep{Bertin2002}, and we degraded the $H$ band to the $K$-band FWHM of 0.43''. We then run \textsc{SExtractor} in dual mode on the two images, to detect and measure all sources in the $Ks$ band, and measure their flux in the H$160$ band (the parameters were optimized to favour the detection of low surface brightness or extended sources). 
We finally selected the objects with signal-to-noise ratio (SN) higher than 5 which were not already present in G13, by means of a cross-correlation with a 1'' search radius (roughly equivalent to 5 $H$ FWHMs, a conservative choice to avoid the risk of including dubious sources). With this procedure, we singled out 184 potential new sources (among these, 5 are not visible at all in the $H$ band (<1 $\sigma$) and 130 are below the $5\sigma$ limit of G13; the remaining 49 are low surface brightness sources). 

\item \textbf{Method 2 - K-band residual image with \textsc{t-phot}}.
While Method 1 is tailored to detect isolated sources, it may fail to detect those that are close to the known $H$-detected sources, because of the rejection within 1''. For such objects we have exploited a second, complementary approach, using the residual image generated by \textsc{t-phot} when used to fit the $Ks$ image with $H$ band priors: the sources that are not included in the $H$ band detection list, and therefore have not been fitted, appear as bright spots. However, we must take into account that bright extended sources typically create irregular residuals, with negative and positive areas that could be mistaken for real sources by a detection algorithm. To cope with this, we created an enhanced RMS map by summing the collage of the fitting models outputted by \textsc{t-phot} as a diagnostic image to the original RMS of the $Ks$ image; this map was then fed to \textsc{SExtractor} for the detection run, thus attributing a lower weight to areas occupied by known $H$-detected sources. We found that this simple technique yielded better results than using the original RMS map, considerably reducing false positives - albeit it may also cause the exclusion of a few potentially true objects very close to bright $H$-detected sources, for which it would be difficult to obtain reliable photometric measurements anyway. 

With this method we detected 267 sources, of which 170 with SN$_{K}$>5; 82 of these are also detected with Method 1. Apart from some spurious detections close to the borders of the image, and despite the effect of weighting mask, we still found many detections clustered around very bright sources, which must be considered as false positives as well. 

\end{itemize}

\begin{figure}
\centering
\includegraphics[width=8.5cm]{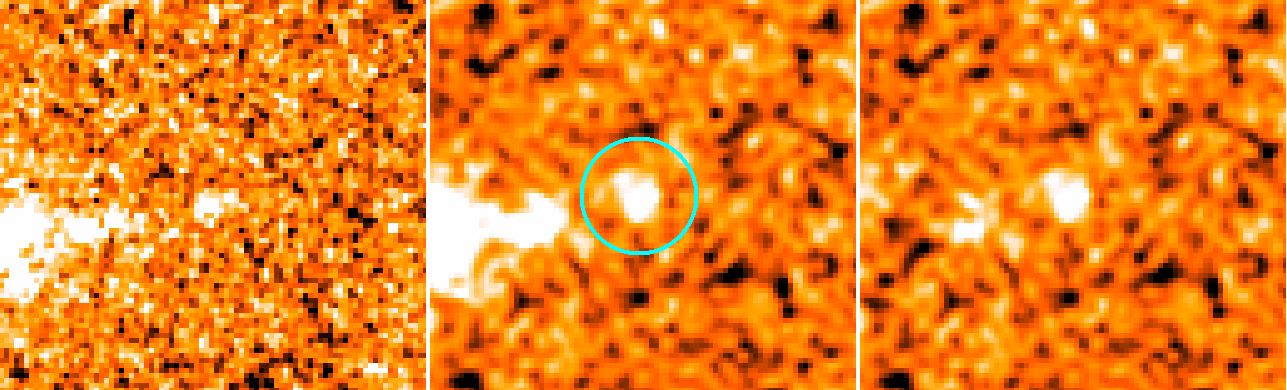}
\vspace{0.25cm}
\includegraphics[width=8.5cm]{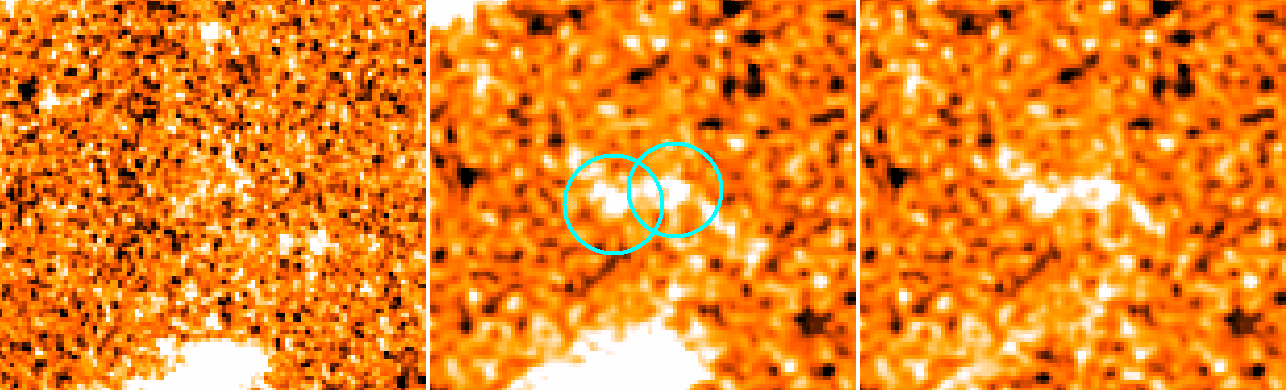}
\includegraphics[width=8.5cm]{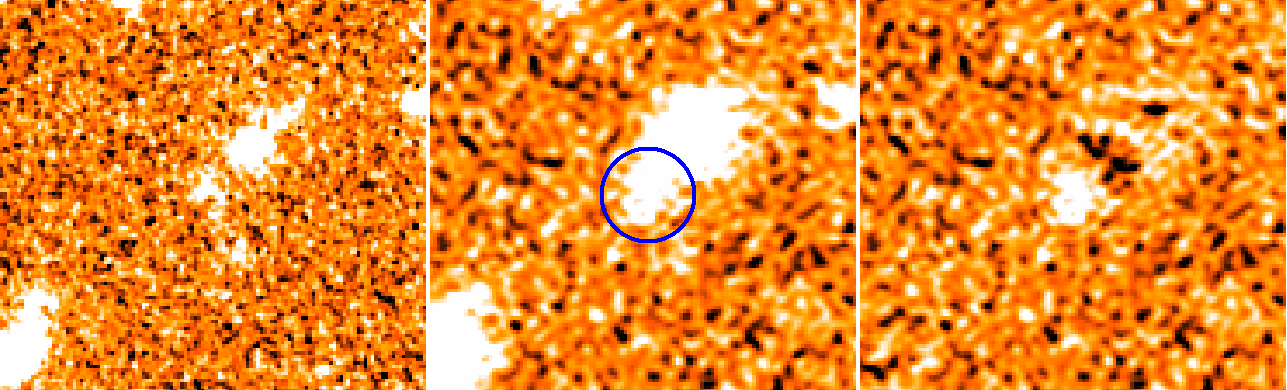}
\includegraphics[width=8.5cm]{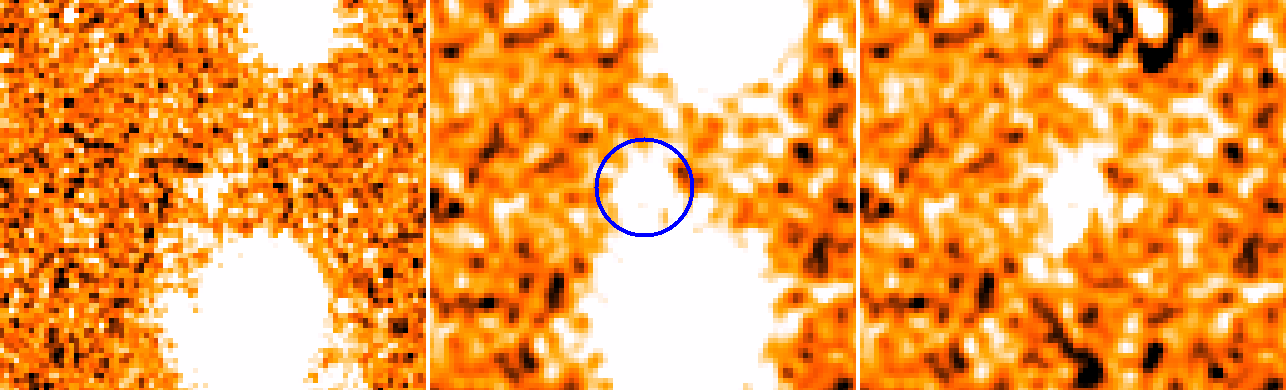}
\caption{Examples of new $K$-detected sources. Left to right: WFC3 $H160$, HAWK-I $Ks$, \textsc{t-phot} residual on $Ks$. Upper panels (cyan circles): examples of sources detected with Method 1 (\textsc{SExtractor} in dual mode and cross-matching with G13). Lower panels (blue circles): examples of sources detected with Method 2 ($K$-band residual image with \textsc{t-phot}). See text for details.}
\label{kdetex}
\end{figure}

\begin{figure*}
\centering
\includegraphics[width=18cm]{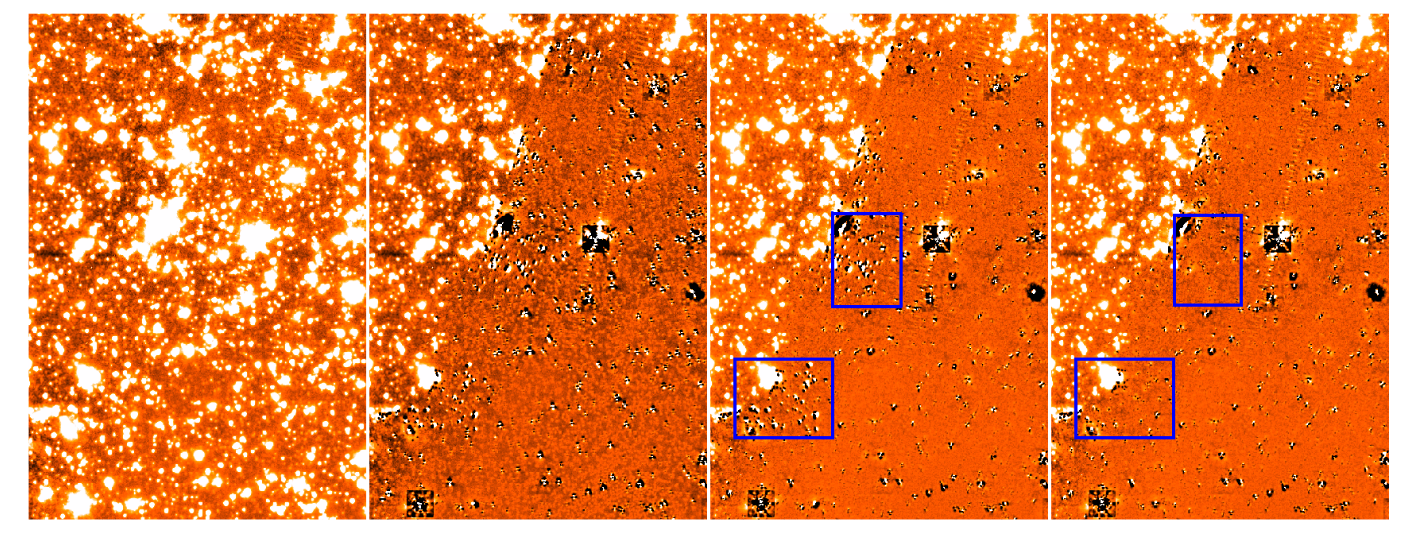}
\caption{From left to right: a portion of the IRAC 3.6$\mu$m mosaic by R. McLure; residuals after \textsc{t-phot} standard fitting using a single convolution kernel; residuals after fitting using a different individual kernel for each source, tailored on the basis of the positional angles of the pointings used to build the mosaic, and with the \textsc{t-phot} global background subtraction option switched on; residuals for the final run, where the \textsc{t-phot} individual kernel registration option is also switched on (in the last two panels, blue boxes highlight regions where the improvement using this technique is evident). See text for more details on the methods.}
\label{comp_irac}
\end{figure*}

\begin{figure}
\centering
\includegraphics[width=8cm]{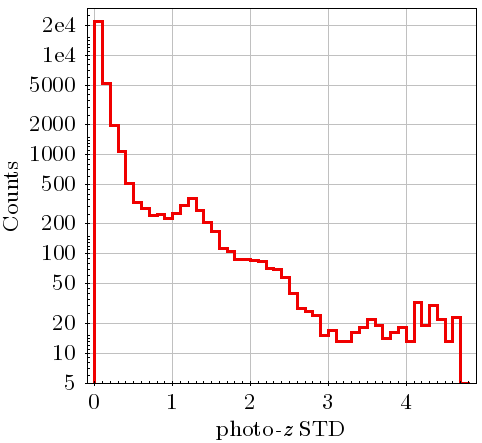}
\caption{Distribution of the standard deviation among the three photometric redshift estimates yielded for each object by \textsc{EAzY}, \textsc{LePhare} and \textsc{z-phot}. For most of the objects the standard deviation is close to zero, indicating good agreement between the three codes. The values are included in the released catalogue, as an indicator of the quality of the median value adopted as the final redshift estimate.}
\label{stdmedian}
\end{figure}

\begin{figure*}
\centering
\includegraphics[width=9cm]{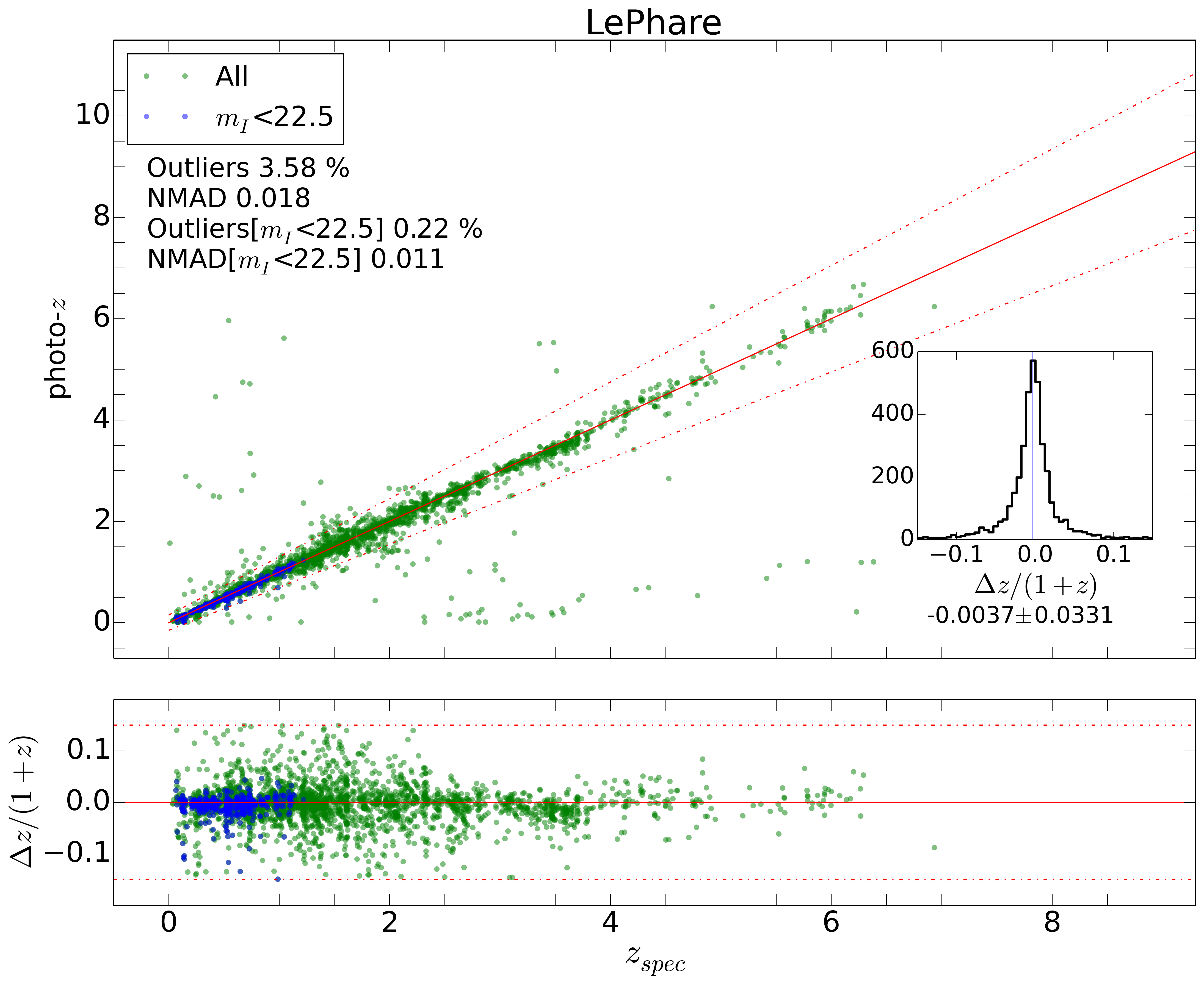}
\includegraphics[width=9cm]{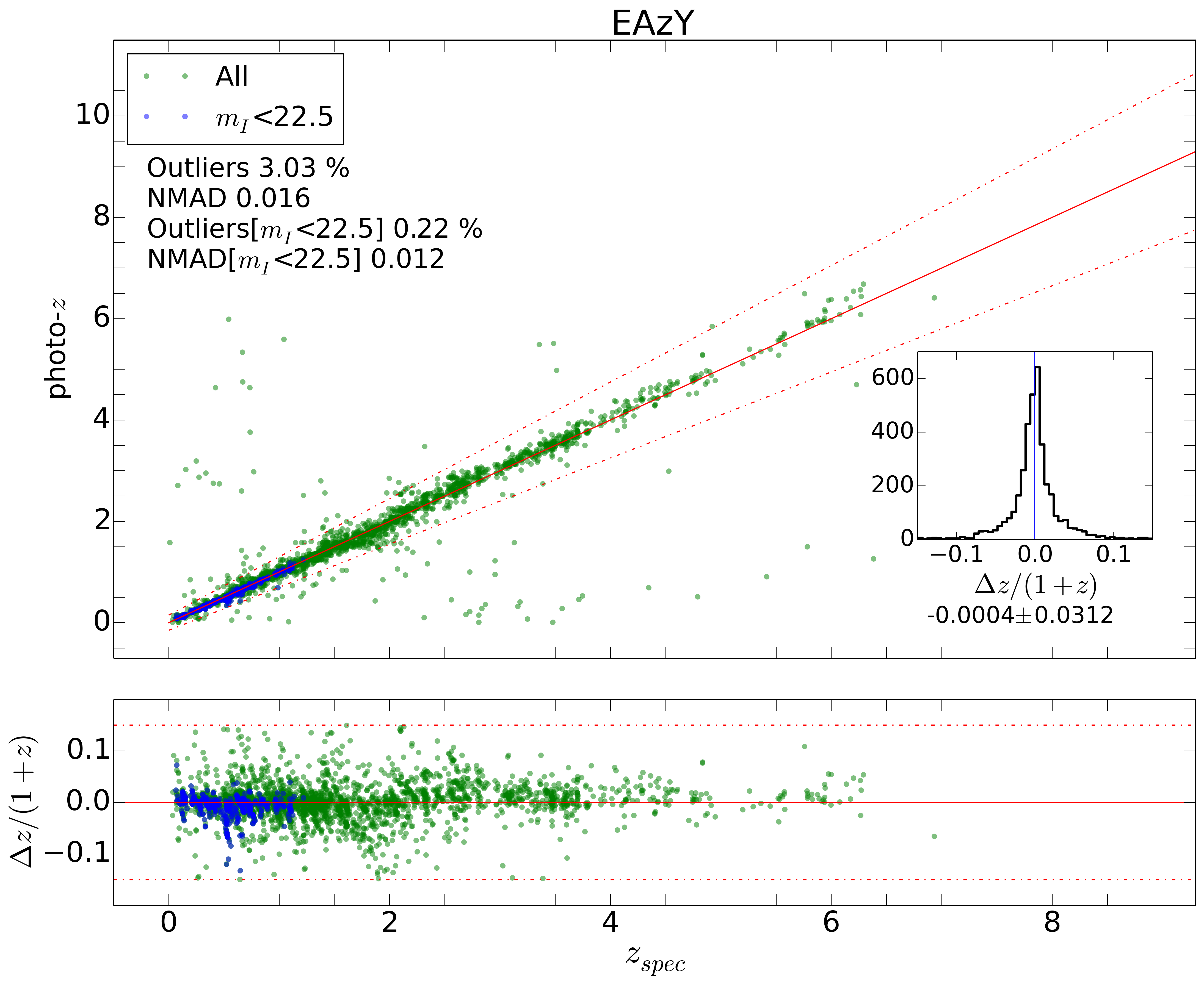}
\includegraphics[width=9cm]{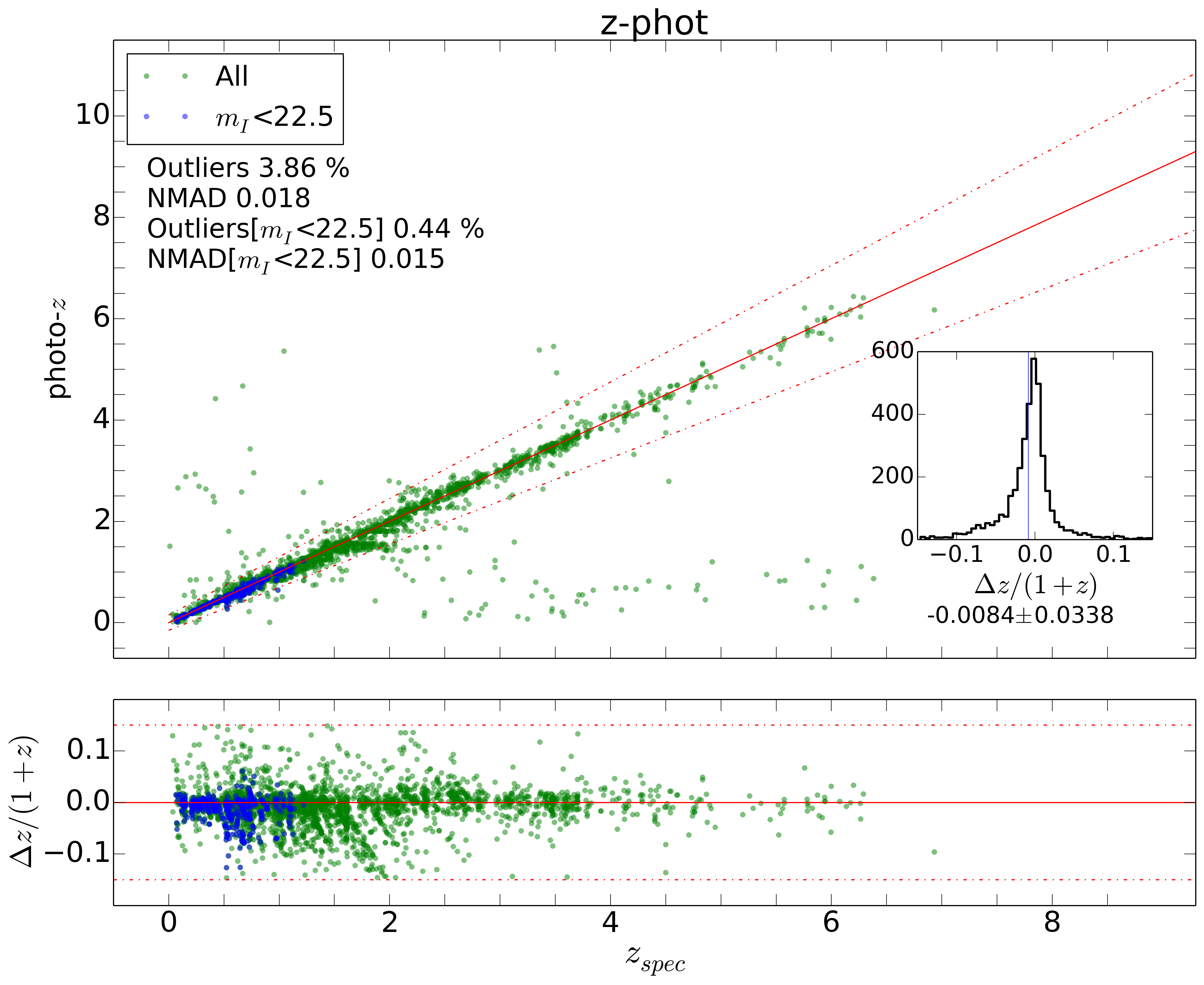}
\includegraphics[width=9cm]{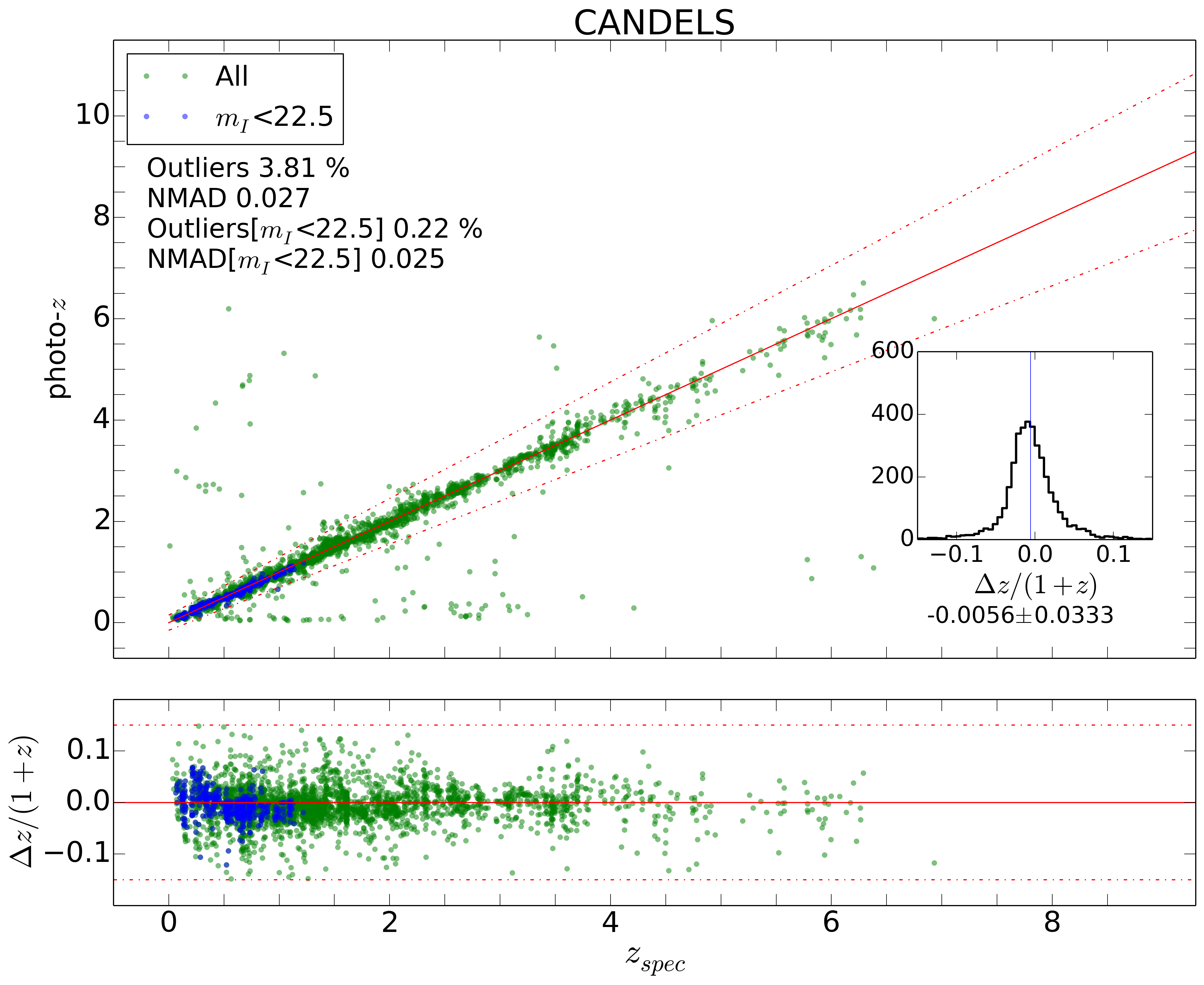}
\caption{Comparison of photo-$z$'s and spec-$z$'s for the runs of (left to right, top to bottom) the three codes \textsc{LePHARE}, \textsc{EAzY}, \textsc{z-phot}, plus (bottom right panel) the CANDELS official estimates from  \citet{Dahlen2013}. Green points are the full sample of 3931 spectroscopic sources described in Sect. \ref{zpc}, while the blue points are the bright tail with $I814$<22.5. In each case, the top sub-panel directly shows photo-$z$ vs. spec-$z$, while the bottom sub-panel shows the corresponding $\Delta z/(1+z_{spec})$ distribution; the small inner sub-panels in the top right corner also show the same quantity as a histogram, with the values of median and standard deviation. See text for more details.} \label{photoz_specz1}
\end{figure*}

To compile the final list of $K$-selected sources we reviewed all the new candidates with SN$_K > 5\sigma$ (that is, the 184 obtained through Method 1 plus the $170-82=88$ additional ones obtained through Method 2), 
and we considered as valid sources only the ones being visible by eye inspection in multiple bands (i.e. $Ks$ plus IRAC, and/or some $HST$ band), or at least having a very solid detection in $Ks$ band (i.e. non spurious with high confidence). By this process, after rejecting 99 sources based on our visual inspection we finally obtained a total of 173 $K$-selected candidates: 75 are detected with both methods, 60 with Method 1 only, and 38 with Method 2 only. A few examples are shown in Fig. \ref{kdetex}. 

\subsection{IRAC-detected sources} \label{iracdet}

Finally, we include in ASTRODEEP-GS43 a list of 5 IRAC-detected sources. We started from the list of 10 $H$-dropouts singled out by W1; in brief, they used the \citet{Ashby2013} 3.6 and 4.5 $\mu$m catalogue for the SEDS survey, and selected $H$-dropouts by means of a cross-correlation with G13, excluding those with an $H$ band counterpart within 2'' (see the original paper for more details). Cross-correlating the coordinates, we found that 7 out of 10 of the W16 galaxies were already present in our list of $Ks$-detected sources; therefore, only the remaining 3 W16 objects were added to our catalogue. Furthermore, we also included two more galaxies found by Wang and collaborators in the same study (priv. comm.), which were excluded from their final published list due to proximity with $H$-detected contaminants. However, since their method was quite conservative and visual inspection ensures they are real $H$-dropouts, we decided to include them in our catalogue.

\section{Photometry} \label{photometry}

We kept the photometric measurements from G13 for the 5 $HST$ WFC3 bands. On the contrary, we measured photometry on the new $HST$ ACS mosaics, with the same procedure adopted in G13 \citep[see also][]{Galametz2013}. We used accurate PSFs from bright, unsaturated stars to build matching kernels between each band and the detection band $H160$, which has the widest FWHM. Isophotal fluxes were then measured on each PSF-matched ACS image using \textsc{SExtractor} in dual-image mode. The total fluxes were then obtained by correcting the detection band Kron flux (i.e. \textsc{SExtractor} \texttt{MAG\_AUTO} in $H160$) with a color term computed as the ratio between ACS and $H160$ isophotal fluxes.

On all the ground based images, including the newly added medium bands, we used the template-fitting software \textsc{t-phot} \cite{Merlin2015,Merlin2016a}. In brief, the code exploits priors cut from high-resolution images to build low-resolution templates; the latter are then used to minimize the difference between a model created as a collage of them, and the real low resolution image. All of the fits were performed using the entire images at once, to ensure that the contamination from neighbors was taken into account. 
Rather than just using priors built only from the $H$ band image (as in the standard practice for $K$ and \textit{Spitzer} bands), for each measurement band we took as priors the cut-outs from the closest $HST$ band in terms of wavelength, provided the SN of the object was higher than 3 in that band (if not, we reverted to the $H$ band prior). We checked that with this approach we obtained cleaner residuals.

Finally, we also used \textsc{t-phot} to measure photometry on all \textit{Spitzer} bands (in this case using $H$ band priors). For these runs, we took advantage of some options of the v2.0 of the code:
\begin{itemize}
    \item each source was fitted with an individual convolution kernel obtained from the local PSF. We built each individual PSF by stacking instrumental PSFs stamps, rotated according to the position angle of each single-epoch observation, and weighted by its exposure time;
    \item a constant background was fitted and subtracted during the fitting process, along with the individual sources fitting;
    \item individual positional registration of the sources was performed after a first fit, to account for astrometric inaccuracies, re-centering the templates to minimize any spurious offset during a second fitting run.
\end{itemize}
These techniques are described in details in \citet{Merlin2016b}. A visual depiction of the improvements obtained by using them is shown in Fig. \ref{comp_irac}.


Concerning the additional 173 $K$-detected sources, we processed all available $HST$ bands using the same technique adopted for the $H$-detected sources, i.e. smoothing them to the widest FWHM, in this case the one of the $Ks$ band. We then run \textsc{SExtractor} in dual mode using the $K$-band image as the detection image for the 60+75 sources detected with Method 1, and the \textsc{t-phot} residuals in the $Ks$ band for the remaining 38 sources detected with Method 2. In this case, considering the faintness of the sources, to obtain the total flux in each band we calculated the color between the measurement band and $Ks$ in a circular aperture of 2 FWHM \citep[a good compromise to avoid strong contamination while retaining the largest fraction of flux, see e.g.][]{Castellano2010}, and applying it to the $K$-band total flux.

The IRAC photometry was again obtained with \textsc{t-phot}; since the sources are typically small and faint, we adopted the PSF-fitting option, directly using the IRAC PSFs as priors rather than exploiting high-resolution cut-outs. The same approach was used for the additional 5 IRAC-detected sources. 

%



The final fluxes are consistent with the previously published ones, within the limits of the different methods used. The figures in Appendix \ref{append} show the quantitative comparisons of the new photometric measurements used in this work with the previous available ones, in a number of reference band-passes. Overall the agreement is always reasonable, but many second order differences can be spotted. In particular, it is already known that the 3D-HST fluxes differ from the CANDELS one \citep[e.g. ][]{Skelton2014, Stefanon2017}, and they retain this bias also when compared to our new photometry. We do not investigate further the reasons of such discrepancies.

\section{Redshifts and physical properties} \label{redshifts}

\begin{figure*}
\centering
\includegraphics[width=14cm]{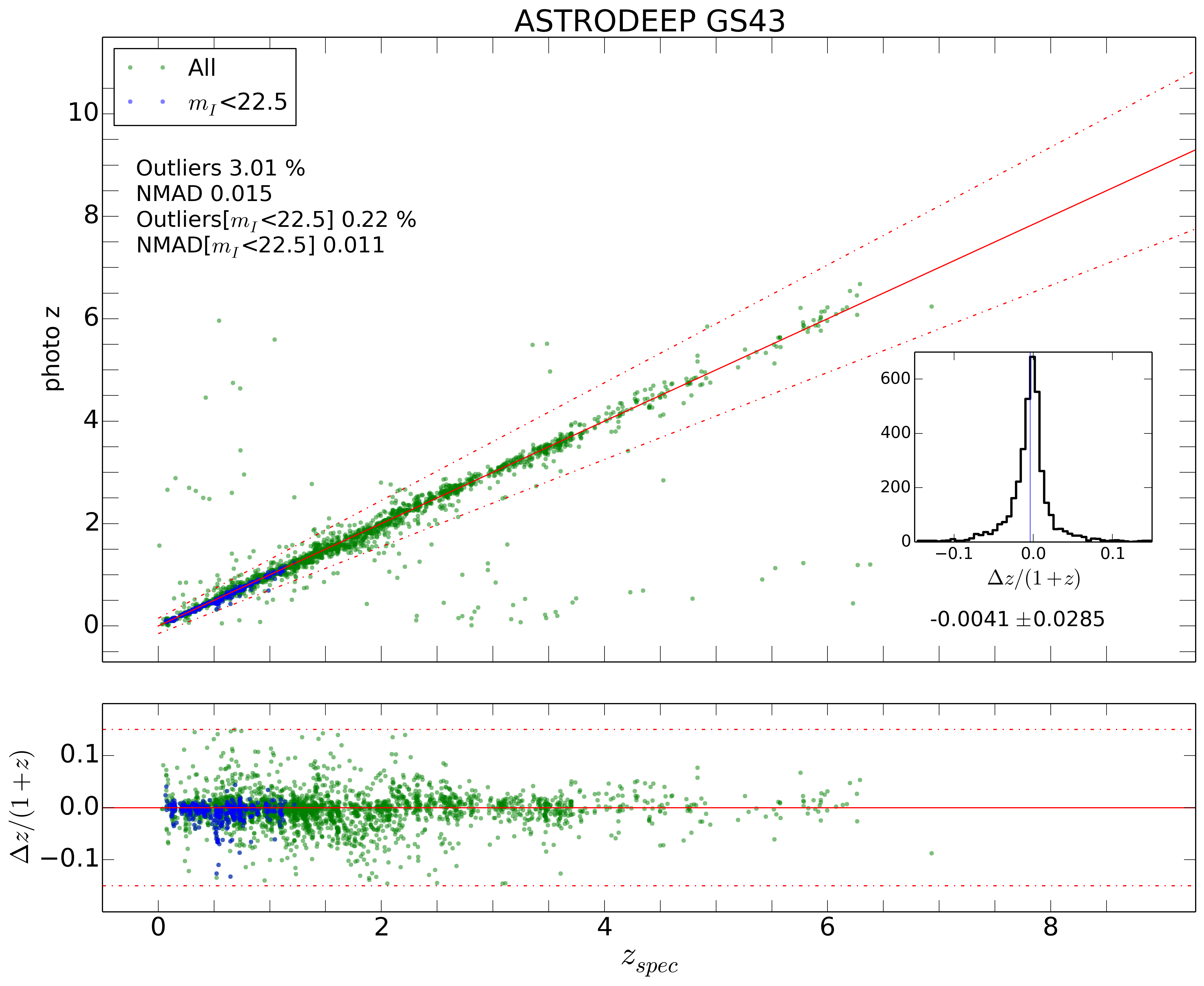}
\caption{Comparison of photo-$z$'s and spec-$z$'s for the best estimate obtained as the median between the three single runs. Green points are the full sample of 3931 spectroscopic sources described in Sect. \ref{zpc}, while the blue points are the bright tail with $I814$<22.5. The top sub-panel directly shows photo-$z$ vs. spec-$z$, while the bottom sub-panel shows the corresponding $\Delta z/(1+z_{spec})$ distribution; the small inner sub-panel in the top right corner also show the same quantity as a histogram, with the values of median and standard deviation.}
\label{photoz_specz2}
\end{figure*}

\begin{figure}
\centering
\includegraphics[width=9cm,height=6cm]{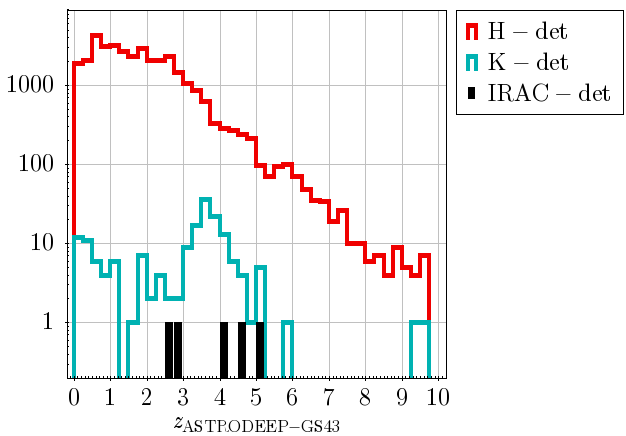}
\caption{Redshift distribution of the three samples in this work: the G13 CANDELS $H$-detected catalogue (red), the 173 $K$-detected sources (cyan), and the 5 additional IRAC-detected sources from \citet{Wang2016} (black).}
\label{zdistr}
\end{figure}

\begin{figure}
\centering
\includegraphics[width=9cm]{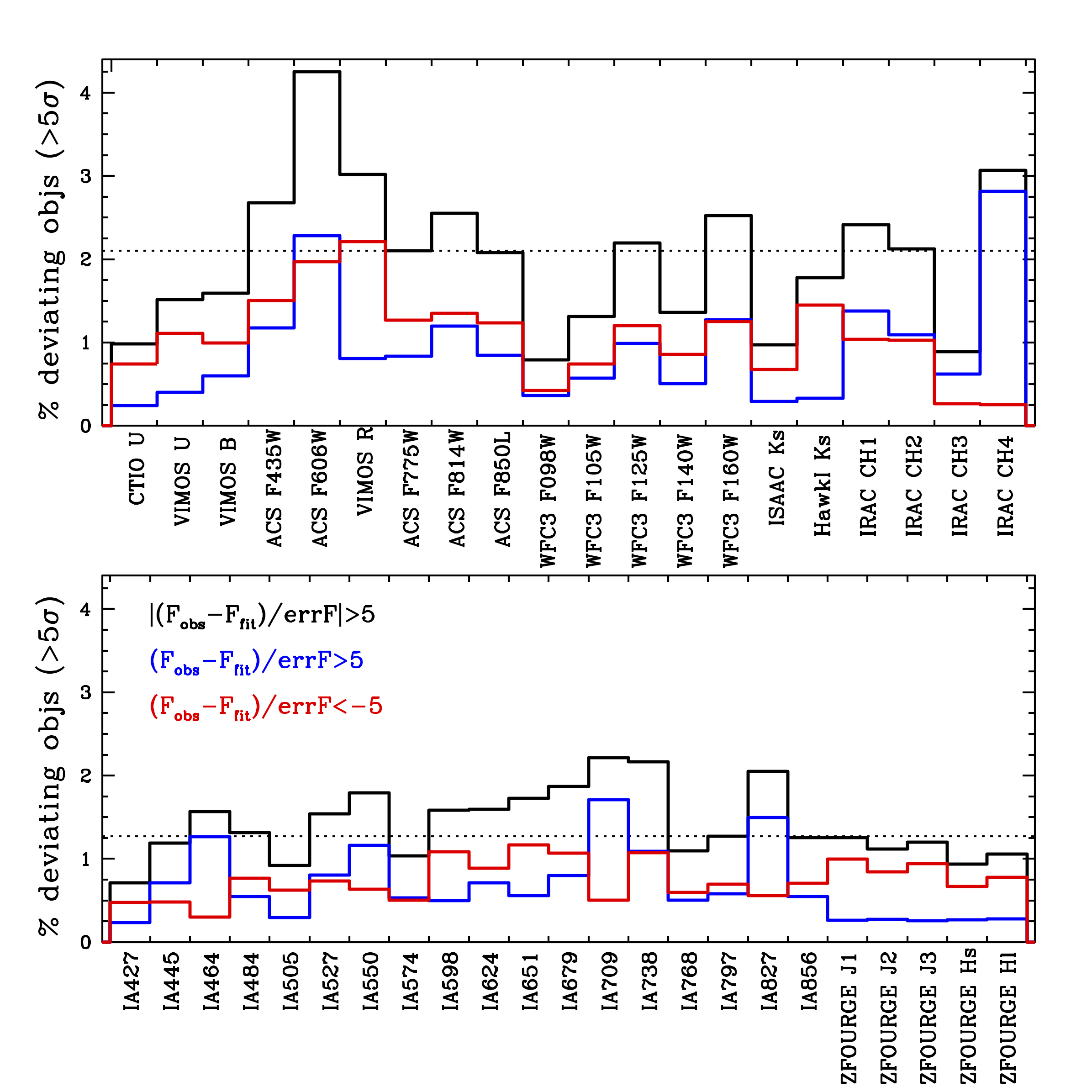}
\caption{Percentage of objects whose photo-$z$ best-fit flux in the \textsc{z-phot} run for the estimation of the physical parameters deviates by more than 5$\sigma$ from the observed one, in any given band. Shown is the result for the $\tau$ models run; the results for the delayed $\tau$ models are very similar. The upper and lower panels show the wide and medium bands, respectively. The total fraction of deviating objects is given by the black histogram, while the blue and red histograms show objects whose best-fit underestimates and overestimates, respectively, the observed flux. The dotted horizontal line is the median value for the considered set of pass-bands (for the totals, these are 2.1\% for the wide bands, and 1.3\% for the medium bands).}
\label{flags3}
\end{figure}

\begin{figure*}[h!]
\centering
\includegraphics[width=8cm]{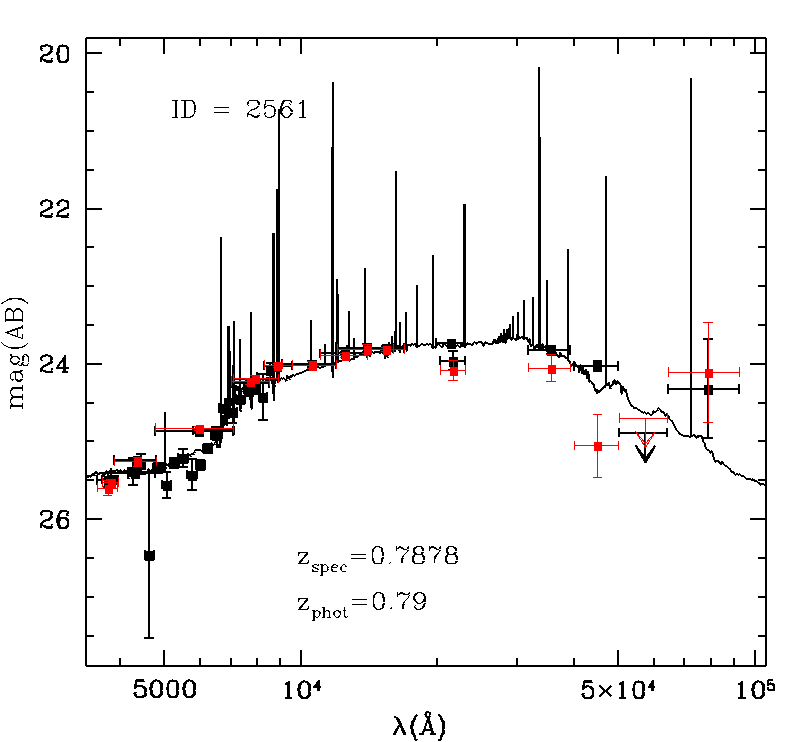}
\includegraphics[width=8cm]{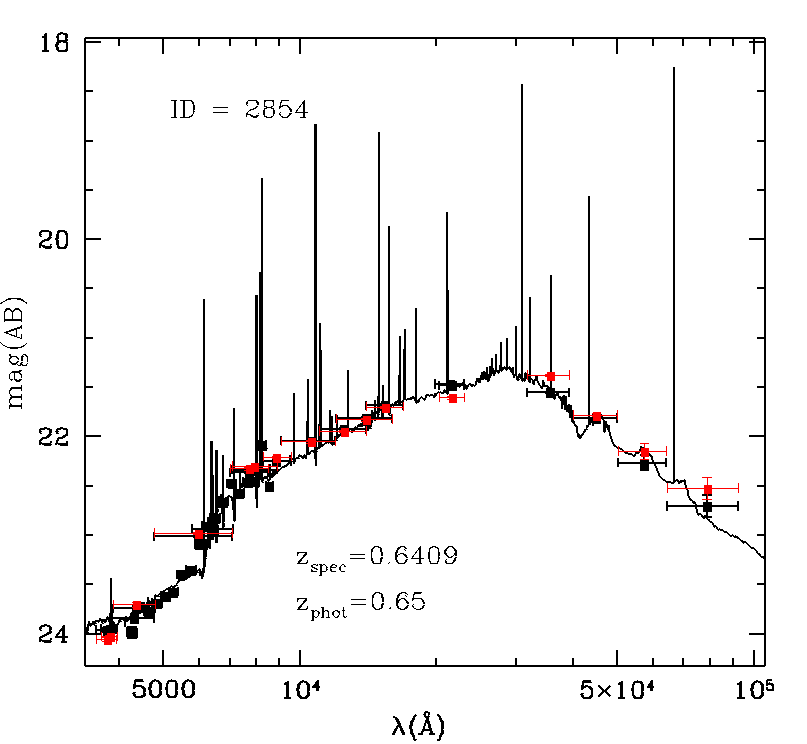}
\includegraphics[width=8cm]{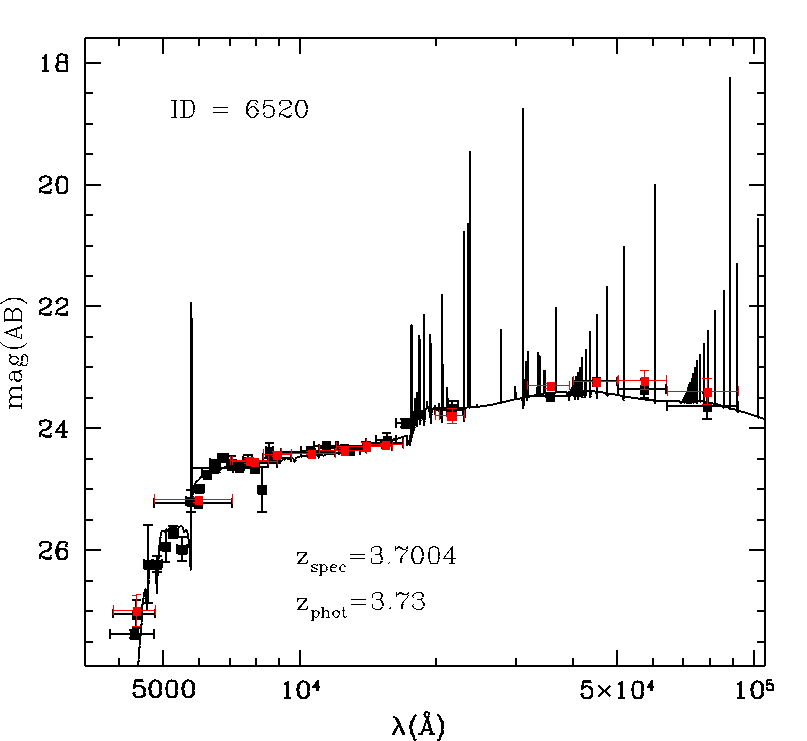}
\includegraphics[width=8cm]{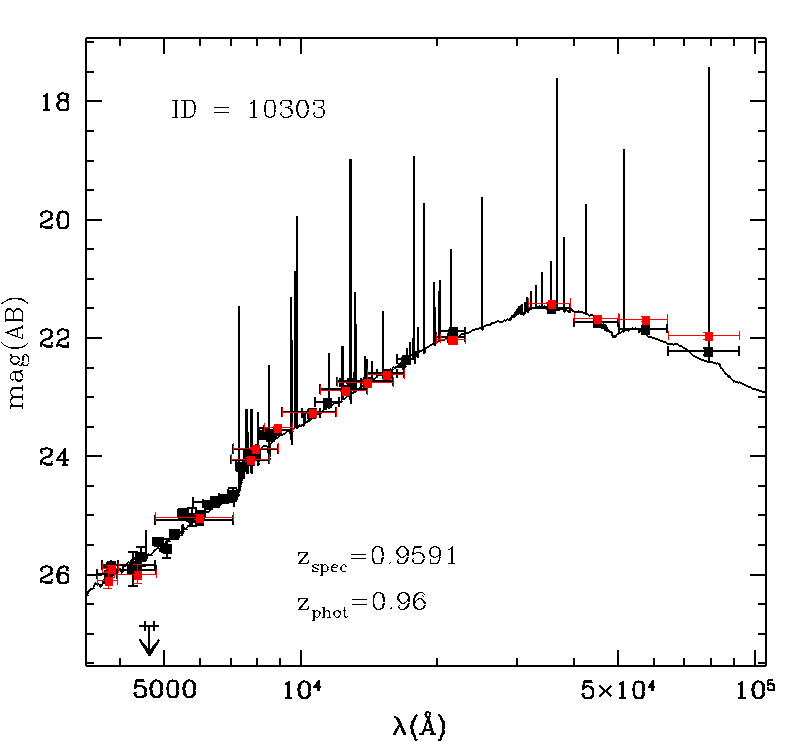}
\includegraphics[width=8cm]{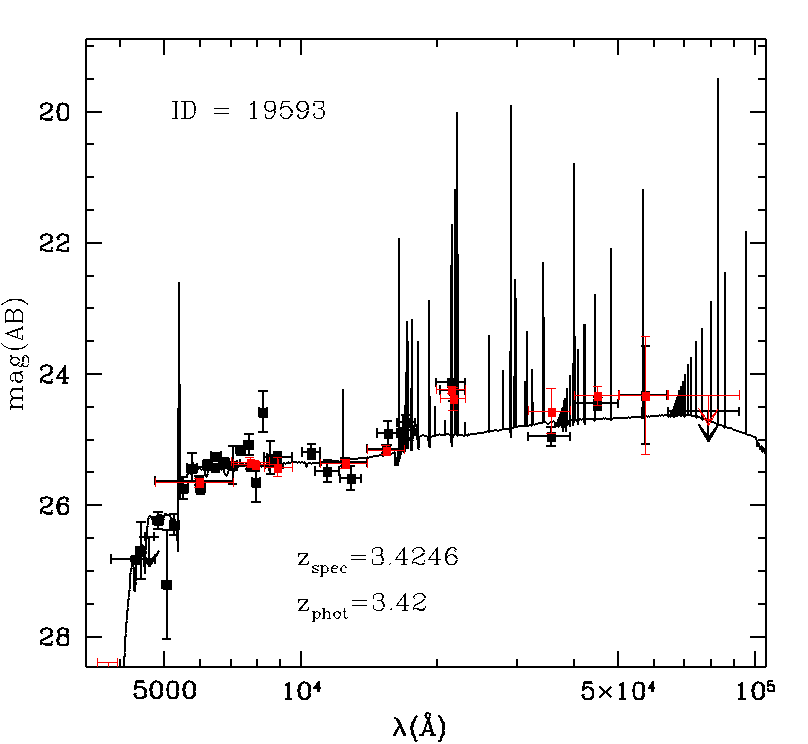}
\includegraphics[width=8cm]{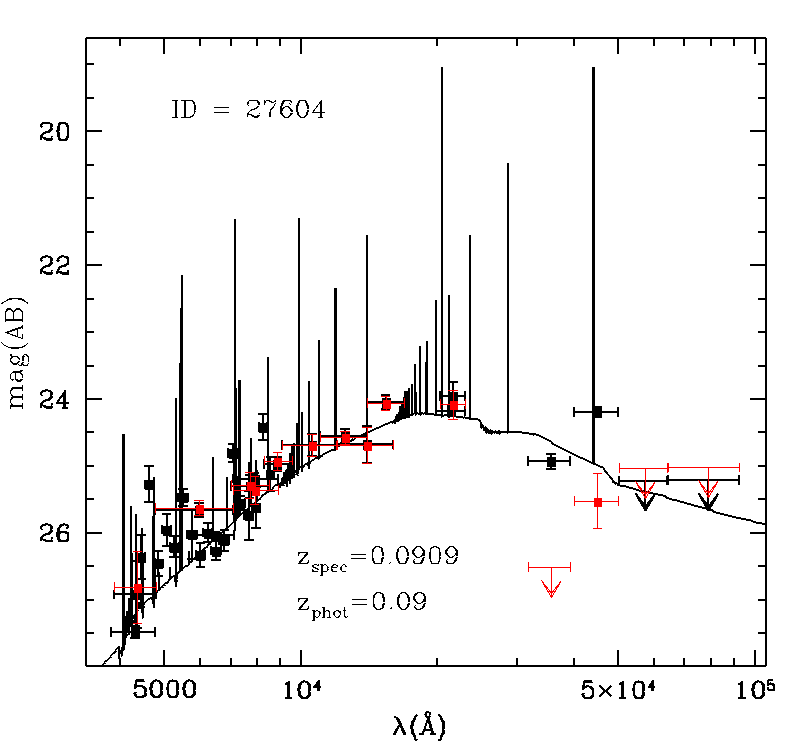}
\caption{Six examples of SED-fitting of photometric sources exploiting the full 43 bands dataset. Red squares are the CANDELS photometry from G13; black squares are the new photometric measurements, and the solid line is the best fit model for the redshift estimate. In these cases, the improved photometric coverage leads to enhanced accuracy in the fit and consequently in the photo-$z$ estimates (reported in the plots) with respect to the CANDELS ones. Globally, this yields an overall improvement in the accuracy of the photometric redshifts, as discussed in Sect. \ref{comp}.}
\label{seds}
\end{figure*}

\begin{figure}
\centering
\includegraphics[width=8.5cm]{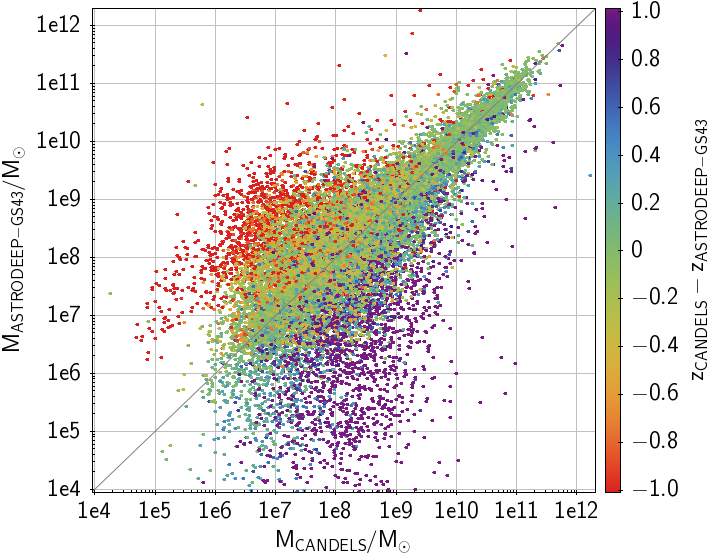}
\caption{Comparison of masses obtained with the \textsc{z-phot} code in the 4 $\tau$ models runs for this work and in the official CANDELS catalogue. The IMF is Chabrier (a factor 1/1.75 has been applied to the new masses, since the IMF in the new runs was assumed Salpeter), the SFH is from $\tau$ models. The color code is proportional to the difference in photo-$z$.}
\label{masses}
\end{figure}

\begin{figure}
\centering
\includegraphics[width=8.5cm]{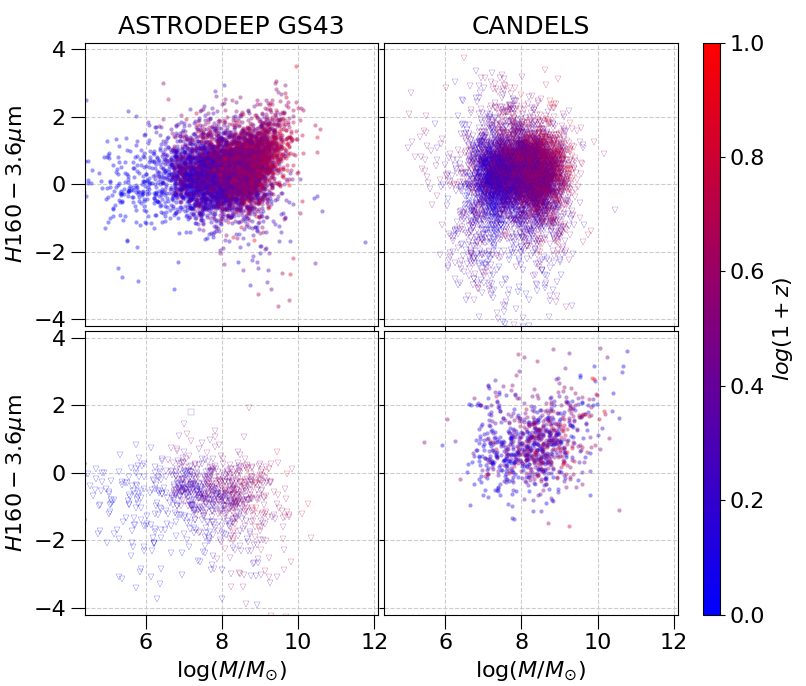}
\caption{Compared color-mass diagrams ($H160-3.6 \mu$m vs. stellar mass) of the GS43 and CANDELS catalogues, color-coded as a function of redshift. In the left panels, stellar mass and color are from the GS43 catalogue; in the right panels they are from CANDELS, photometry from G13 and masses from \citet{Santini2015}; in both cases we considered the standard $\tau$ models including emission lines. In the upper panels we show the sources which in GS43 are detected (SN>1) in IRAC CH1 and are upper limits in G13; the lower panels show the opposite.}
\label{cm}
\end{figure}

In this Section we describe how we obtained the estimates of the photometric redshifts and of the main physical parameters, via spectral energy distribution (SED) fitting, for all the non-spectroscopic G13 $H$-detected sources and for the additional 178 $K$ and IRAC-detected ones. Note that the released catalogue lists the ``best'' redshift estimate, which is the spectroscopic one whenever available and of good quality, and the photometric one otherwise.


\subsection{Available spectroscopic data} \label{specz}

First of all, we compiled a list of 4951 high quality and publicly available spectroscopic redshifts (of which 4829 are of galaxies and 122 of stars) from a number of surveys and references (the full list is included in the catalogue \texttt{README} file). When two or more measurements were available for the same object, the one with the highest quality flag assigned was kept. We then used the spec-$z$'s to optimise the calibration of photometric redshifts as described in Sections \ref{zpc} and \ref{comp}. We point out that the list used in this procedure includes the VANDELS release DR3 that was available at the time of submission; however, during the revision of this work the final release DR4 has become available \citep{Garilli2021}. After checking that differences were not substantial, we have not repeated the optimisation procedure, but we have instead used DR4 in the final \textsc{z-phot} runs to determine the physical properties of the sources (Section \ref{physpar}), and in the published catalogue.

\subsection{Star/galaxy separation} \label{sgsep}

Before proceeding to estimate the photometric redshift and physical parameters of the catalogued galaxies, we cleaned the list of detected sources flagging out those which can be safely identified as stars. First of all, we identified and removed the spectroscopic ones; then we proceeded as in \citet{Grazian2007}, combining the \texttt{CLASS\_STAR} estimator (outputted by \textsc{SExtractor} from the original G13 detection run on the $H$ band) with the analysis of the $BzK$ diagnostic plane \citep{Daddi2004}, which we performed exploiting our new photometry from $B435$, $Z850$ and Hawk-I $Ks$. In practice, first we selected the objects brighter than $H=24.5$ (this is a conservative cut, but given that fainter magnitudes are prone to large photometric errors, and that in the low brightness regime high-redshift galaxies dominate over faint stars, we preferred the risk of wrongly classifying a real star as a galaxy, rather than excluding real galaxies misjudging them as stars). Then, among this selection we flagged as stars the objects having \texttt{CLASS\_STAR}>0.85 \emph{and} satisfying the $BzK$ selection criterion $(z-K)<0.3\times(B-z)-0.5$. Combining the spectroscopic list with the one obtained with this technique we end up with a final list of 174 stars, which we include in the catalogue, but that are not part of the subsequent analysis.

\subsection{$H$-detected galaxies} \label{photoz}

To obtain photometric redshifts for the $H$-detected sources in G13 that do not have a spectroscopic observation, we exploited independent estimates from three SED-fitting software tools fed with the new photometric catalogue, namely \textsc{LePhare} \citep{Arnouts1999,Ilbert2006}, \textsc{EAzY} \citep{Brammer2008},  
and \textsc{zphot} \citep{Fontana2000}. We run \textsc{EazY} and \textsc{zphot} on our local machines, while \textsc{LePhare} was run remotely via the \textsc{GAZPAR} web portal\footnote{\textit{https://gazpar.lam.fr/home}}. 

For \textsc{LePhare} we used the same setting and parameters used in \citet{Ilbert2009} for their COSMOS catalogue, which include templates by \citet{Polletta2007} plus additional starburst templates generated using BC03.

For the \textsc{EAzY} runs we used the built-in set of templates described in \citet{Brammer2008}, Sect. 2.2; we did not apply the bayesian prioring option, because testing the possible configurations with a preliminar photo-$z$ vs. $z_{spec}$ comparison we found that including the priors yielded very similar results in terms of absolute dispersion, but also a slightly larger number of outliers, in particular a few of objects at 1<$z_{spec}$<3 wrongly estimated to have $z\simeq0$. Given that the priors tend to disfavour high redshift solutions, which in turn should be the most likely ones in deep surveys like CANDELS, we preferred to proceed without them.

Finally, for the \textsc{zphot} runs we compiled a library with two star formation histories (SFHs): a standard exponentially declining ``$\tau$ model'' in which the star formation rate is $SFR(t) \propto  exp[-(t-t_0)/\tau]$ (where $t_0$ is the beginning of the star formation activity); and a ``delayed-$\tau$'' SFH for which $SFR(t) \propto (t^2/\tau) \times exp[-(t-t_0)/\tau]$. We used \citet[][BC03]{Bruzual2003} templates including nebular emission lines following \citet{Castellano2014} and \citet{Schaerer2009}, assuming a \citet{Salpeter1959} IMF, with a standard range of metallicities (0.02, 0.2, 1 and 2.5 $Z/Z_{\odot}$ depending on the age of the models) and of dust extinctions \citep[according to the][law] {Calzetti2000}.



\subsubsection{Zero-point corrections} \label{zpc}

For each of the three runs with different codes, we computed and applied independent zero-point corrections to the measured fluxes, to obtain better final redshift estimates. We determined the corrections as follows. First, we made a run on the original photometric catalogue, after having cleaned it removing unreliable entries (i.e., magnitudes outside a fiducial range, which after some trials we set to $[15,28]$ for all ground-based bands, $[15,30]$ for \textit{HST} bands, and $[15,27]$ for IRAC; and upper limits below some reasonable values, again set by trials as 27 for ground based bands, 28 for $HST$, and 26 for IRAC). Then, we verified the output against a sub-selection of 3931 sources from the full spectroscopic sample described in Sect. \ref{specz}, in the following way: we excluded from the full sample the sources having covariance index $>1$ in one or more \textsc{t-phot} runs (because their photometry is unreliable due to dramatic blending), or having already being identified as AGNs by \citet{Cappelluti2016} or VANDELS DR2 \citep{Pentericci2018, McLure2018}\footnote{\textit{https://www.eso.org/sci/publications/announcements/sciann17139.html}}; also, we excluded sources whose photometric redshift fit had $\chi^2=\sum [(f_{meas}-f_{model})^2/\sigma_{meas}^2]>500$ (we found by trial that this threshold is a good compromise to have robust results, while keeping as many spectroscopic sources as possible). 
We used this reduced spectroscopic sample to compute zero-point corrections for each band: in practice, we run the photometric codes once per band, keeping the redshift fixed at the spectroscopic value and ignoring the considered band in the fit, so that the flux of the best-fit model in that band is not affected by its observed value; the median of the difference between the observed magnitudes and the model magnitudes for all the objects gives the correction for the considered band. We then performed a second run, using the zero-point corrected catalogue\footnote{We applied this procedure directly for \textsc{zphot} and \textsc{EAzY}, while for \textsc{LePhare} we took advantage of the dedicated option on the remote \textsc{GAZPAR} portal}. For most of the bands the resulting corrections have absolute values below $\sim0.05$ mag, for a few exceptions they range from -0.15 (\textit{Subaru} $I827$) to +0.25 (IRAC CH4) mag. This procedure led to a small but consistent improvement in the final statistics of the photometric vs. spectroscopic comparison.

\subsubsection{Final redshift estimates} \label{comp}

We did not attempt a combination of the $p(z)$'s from different codes, like e.g. done by \citet{Dahlen2013}, since it would require specific fine tuning and adaptation to a low number of independent estimates. Instead, we preferred to use a simpler approach similar to the one used in our previous analysis of the Frontier Fields \citep[e.g.][]{Castellano2016,DiCriscienzo2017}, which already enabled a very good quality of the photo-$z$ statistics. We combined the estimates from \textsc{LePhare}, \textsc{EAzY} and \textsc{z-phot} taking the median value of the three; we found that the median reduces both the fraction of outliers $\eta$ (defined as the sources having  $|z_{phot}-z_{spec}|/(1+z_{spec})>0.15$) and the NMAD\footnote{The normalized median absolute deviation is defined as $1.48\times \mbox{median}(|z_{phot}-z_{spec}|/(1+z_{spec}))$.} on the spectroscopic sample with respect to any of the single runs taken alone. If one of the codes fails the fit, the algorithm takes the mean between the other two; we note that we excluded from the averaging process any spurious estimate equal to one of the extremes of the allowed redshift range, i.e. zero or 10. To assess object by object whether the median value is a reasonable choice, in the catalogue we also list the three independent estimates and their standard deviation, whose distribution is shown in Fig. \ref{stdmedian}: for most of the sources ($\sim25000$) it is less than 0.2.


Figs. \ref{photoz_specz1} and \ref{photoz_specz2} summarize the accuracy of the results, showing the values of ASTRODEEP-GS43 best photo-$z$ estimates for the objects in the spec-$z$'s sample used for the zero-point correction calibration. Our final estimates have NMAD=0.015 and $\eta$=3.01\%; considering only the bright objects ($I814<22.5$) we obtain NMAD=0.011 and $\eta$=0.22\%, which is comparable to the values obtained by \citet{Ilbert2013} for their UltraVISTA DR1 30 bands catalogue on COSMOS (with the ``zCOSMOS bright'' sample of $\sim$9400 spectroscopic redshifts at $i^+<22.5$, they find NMAD=0.008 and $\eta$=0.6\%). The first three panels of Fig. \ref{photoz_specz1}, and Fig. \ref{photoz_specz2}, show the comparison of the photometric and spectroscopic redshifts for each of the three independent runs and for the final median average, also reporting the NMAD and $\eta$ statistics. The three runs yield comparable accuracy, with \textsc{EAzY} giving slightly better results on the full sample, and \textsc{LePhare} on the bright tail. Note the percentage of outliers for the bright tail is the same for all codes, due to the fact that only one source is outside the range in all of the three runs.

\subsection{Additional $K$- and IRAC-detected sources}

For the additional $Ks$ and IRAC-detected sources we only used \textsc{zphot} to estimate the redshifts, with the same library of models described in Sect. \ref{photoz}; since no spectroscopic redshifts are available for calibration, we used the same ZP corrections obtained for the $H$ band catalogue.

Fig. \ref{zdistr} displays the distribution of the evaluated redshifts for the three samples in the catalogue (the G13 list, the new 173 $K$-detections, and the 5 IRAC detections).


\subsection{Physical properties} \label{physpar}



We used \textsc{z-phot} to evaluate the physical parameters of all the sources, keeping the redshift fixed to the best estimate obtained as described above, and using the BC03 library, including nebular emission lines. We made two runs, one with standard exponentially declining SFHs ($\tau$ models), and one with delayed-$\tau$ models (see Sect. \ref{photoz}); in the final catalogue we list stellar masses and star formation rates, with the corresponding 1$\sigma$ uncertainties, for both of these runs.

To further the overall quality of the fit and check whether any band has problematic photometry, we used these last \textsc{z-phot} runs to check the fraction of objects whose best-fit flux strongly (i.e., more than 5$\sigma$) deviates from the observed flux, in each of the 43 bands. The results are shown in Fig. \ref{flags3}. For the $\tau$ models run, on average only 2.1\% (1.3\%) of the sources are not well represented by the best-fitting templates within these limits considering the wide (medium) bands, the two bands with the worst performance being ACS $F606W$ and IRAC CH4; even in these two cases, however, the fraction of strongly deviating sources is respectively $\sim$4\% and $\sim$3\% of the total number of catalogue entries. The results for the delayed-$\tau$ are similar.

Finally we assigned a quality flag to each source, for both runs, considering the (non-normalized) $\chi2$ of the fitting processes. We visually inspected the fitted SEDs of a random sample of objects and statistically evaluated the distribution of $\chi2$ values, finding that $\chi2=5$ is a reasonable watershed between acceptable and non satisfying results. We also assigned separate flags to stars, identified as described in Section \ref{sgsep}, and to AGNs, using the catalogue by \citet{Cappelluti2016} together with the VANDELS DR4 data (we release the photo-$z$ and physical parameters estimates anyway for AGNs).
The flags are described in Table \ref{qf} (the fractions are from the delayed-$\tau$ fit, but they are almost identical for the $\tau$ run); with the chosen criteria, $\sim6\%$ of the objects are assigned a bad flag, indicating an unreliable fit. 
\begin{table}[h!]
\renewcommand{\arraystretch}{1.5}
\caption{Quality flags for physical parameters}
\centering
\begin{tabular}{ | l | l | l | l |}
\hline
Flag & Description & $\chi2$ & Fraction \\ \hline
0 & Good/acceptable fit & $\leq$5 & 93.37\%\\ 
1 & Bad fit & >5 & 5.38\% \\
2 & Fit failed & 1.E10 & 0.02\% \\
3 & Star & - & 0.50\% \\
4 & AGN & - & 0.73\%\\
\hline
\end{tabular} \label{qf}
\end{table}

\subsection{Comparison with CANDELS data} 

To conclude our analysis, we compared our best photometric redshift estimates with the official CANDELS provided in the catalogues data release \citep{Dahlen2013}. The results in terms of photo-$z$/spec-$z$ comparison for the CANDELS catalogue are shown in the fourth panel of Fig. \ref{photoz_specz2}. 
The new ASTRODEEP-GS43 redshift estimates reach a higher accuracy both in terms of the NMAD and of the outliers fraction.

In Fig. \ref{seds} we show the SED-fitting of six sources with spectroscopic redshifts, which have improved the photometric redshift estimate with respect to the CANDELS fit. The combination of (i) the higher quality of the images, (ii) the new photometric software, (iii) the finer wavelength coverage (thanks to the addition of the medium bands), and (iv) the method adopted to evaluate the redshifts, yields a high accuracy in the determination of the best-fit model of the source. This allows for an optimal tracing of the underlying spectral features, and therefore of a good photo-$z$ estimate, which in the displayed cases is closer to the spectroscopic value than the CANDELS one. Statistically, these cases are more numerous than the opposite, leading to an overall improvement of the global accuracy of the photo-$z$ estimate, as discussed above.

We also compared our statistics with those by \citet{Kodra2019}, who combined four independent redshifts estimates on the original 17 bands G13 catalogue using the minimum Frechet distance method; we take their \texttt{mFDa4\_weight} estimate, which is the one they choose as ``best'' in absence of spectroscopic values. Also in this case, we find overall better results, their statistics being NMAD=0.027 (0.023 for $m_I<22.5$) and $\eta$=4.46\% (0.22\%). We checked that this result mainly depends on the improved photometric quality and wavelengths coverage, since taking the median of the \textsc{EAzY}, \textsc{LePhare} and \textsc{z-phot} runs used in Kodra's analysis (i.e. the original CANDELS estimates) we find that the statistics are still worse than the ones we get for our new data.

Fig. \ref{masses} shows the comparison between the new stellar masses (shown are the values obtained with the $\tau$ model runs) and the CANDELS ones from \citet[][we used the \texttt{6a\_tau\_NEB} estimate for consistency with the new models]{Santini2015}; we applied a conversion factor of 1.75 to compensate for the different IMF, which is \citet{Chabrier2003} for CANDELS and \citet{Salpeter1955} for ASTRODEEP-GS43. The agreement is good, except of course for the objects having large differences in the estimated redshift in the two catalogues; for the galaxies with $|z_{CANDELS}-z_{GS43}|$<0.1, the mean relative difference in mass is 26.3\%, and the two estimates have distributions which are consistent within the error budget (for 67.7\% of the galaxies the CANDELS value is whithin the 1$\sigma$ confidence interval of the new mass estimate).


Finally, Fig. \ref{cm} shows a color-mass diagram ($H160-3.6 \mu$m vs. stellar mass) of a subsample of the catalogue, and of the corresponding CANDELS catalogue. In the left panels, color and stellar mass are from the GS43 catalogue; in the right panels they are from CANDELS, respectively from G13 and \citet[][we used again the \texttt{6a\_tau\_NEB} estimate]{Santini2015}. In the upper panels we show the sources which are detected (SN>1) in IRAC CH1 in GS43, but are upper limits in G13; the lower panels show the opposite case. Thanks to the combination of the new deeper mosaics and the new improved photometry with \textsc{t-phot}, in ASTRODEEP-GS43 we gain 6369 detections, while losing 878. We note that some sources which had been catalogued as red and massive in CANDELS, having CH1 detections at $>1\sigma$, appear now to be upper limits (this is likely due to the fact that \textsc{t-phot} allowed for a better decontamination from neighboring objects, lowering their flux).


\section{Summary and conclusions} \label{summary}

We have presented ASTRODEEP-GS43, a new photometric catalogue for the GOODS-South field, which includes 43 pass-bands (we release total fluxes and corresponding 1$\sigma$ uncertainties), photometric redshifts for sources without an available spectroscopic estimate, and physical properties (stellar mass and SFR) of 35108 objects: 34930 are the $H$-detected ones from the original CANDELS G13 catalogue, while 173 are additional sources detected in $Ks$, and 5 in IRAC bands (the latter from the W16 study, including 2 that were not present in their original published list).

The new ASTRODEEP-GS43 redshifts and physical parameters estimates prove to be consistent with previous releases; thanks to the use of new, deeper images for \textit{HST} ACS and \textit{Spitzer} IRAC bands, the adoption of new techniques for photometric measurements (\textsc{t-phot} v2.0 with adaptive prioring, background subtraction, individual kernels and astrometric registration), and the combination of different software tools for photometric redshifts estimates (\textsc{LePhare}, \textsc{EAzY} and \textsc{z-phot}),  
comparisons with the CANDELS official data show an overall good agreement, with a noticeable improvement in the quality of the photometric redshift estimates, which reach NMAD=0.015nd $\eta=3$\%; considering only the bright objects ($I814$<22.5) we obtain NMAD=0.011 and $\eta$=0.22\%.

The catalogue is available for download from the \textsc{ASTRODEEP} websiste, at the url  http://www.astrodeep.eu/astrodeep-gs43-catalogue/ .

\begin{acknowledgements}
Part of this study was funded using the EU FP7SPACE project ASTRODEEP (Ref.  No: 312725), supported by the  European Commission.  \\
This work is partly based on tools and data products produced by GAZPAR operated by CeSAM-LAM and IAP.\\
This research has made use of the SVO Filter Profile Service (http://svo2.cab.inta-csic.es/theory/fps/) supported from the Spanish MINECO through grant AYA2017-84089.\\
The Cosmic Dawn Center is funded by the Danish National Research Foundation under grant no.\ 140. BMJ is supported in part by Independent Research Fund Denmark grant DFF - 7014-00017.
The authors thank Tao Wang and Mengyuan Xiao for their contributions to the work.
\end{acknowledgements}

\bibliographystyle{aa}
\bibliography{biblio} 

\onecolumn

\appendix
\section{Photometric comparisons with other catalogues} \label{append}
\begin{figure}[hbt!]
\centering
\includegraphics[width=15cm]{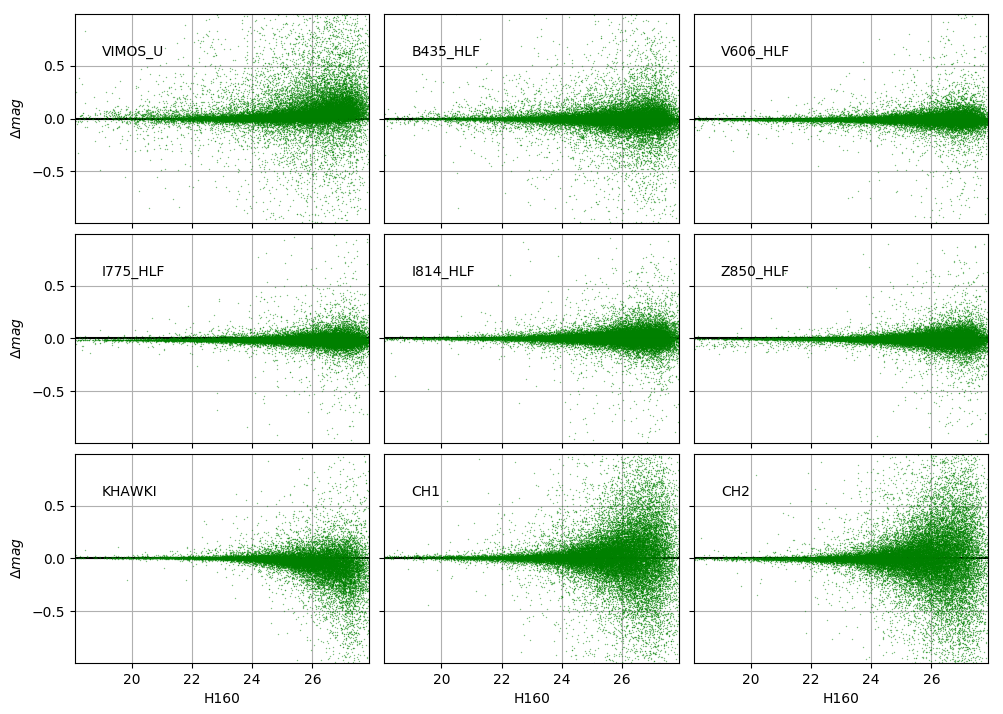}
\caption{Comparison of measured magnitudes between this work and the original CANDELS G13 catalog, for 9 reference bands.}
\label{comp_candels}
\end{figure}
\begin{figure}[hbt!]
\centering
\includegraphics[width=14.5cm]{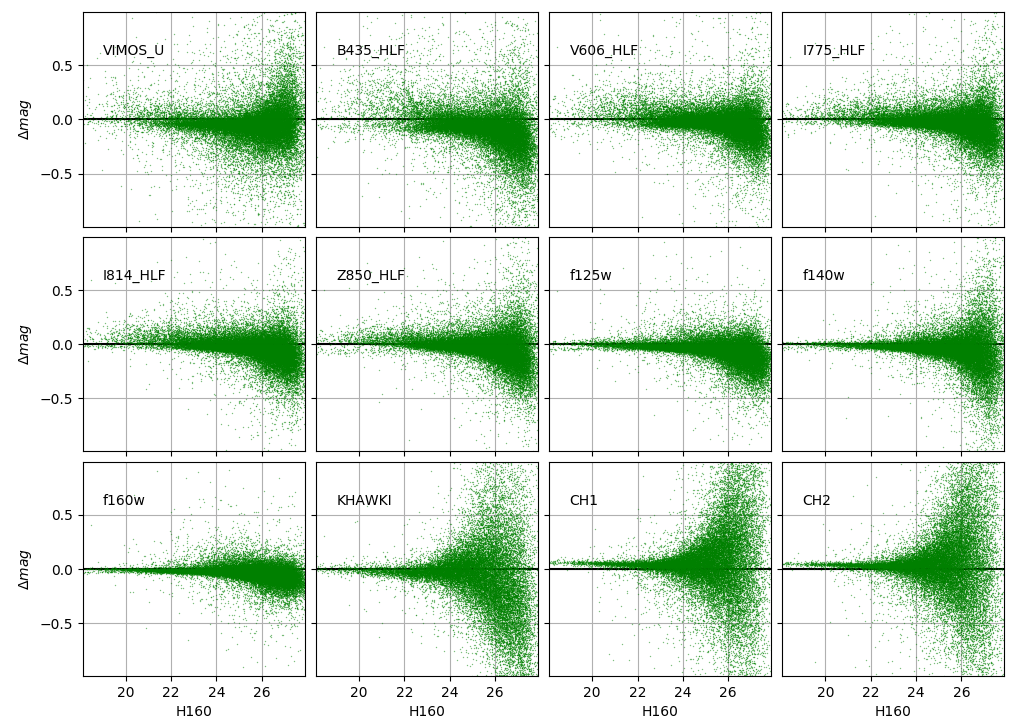}
\caption{Comparison of measured magnitudes between this work and the 3D-HST catalog, for 12 reference bands.}
\label{comp_3dhst}
\end{figure}
\begin{figure}[hbt!]
\centering
\includegraphics[width=14.5cm]{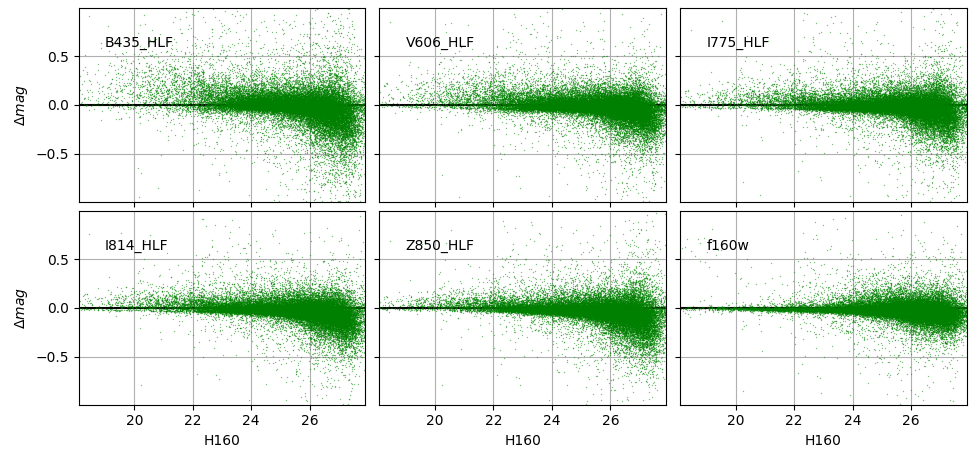}
\caption{Comparison of measured magnitudes between this work and the HLF catalog, for 6 reference bands (ACS).}
\label{comp_hlf}
\end{figure}
\begin{figure}[hbt!]
\centering
\includegraphics[width=14.5cm]{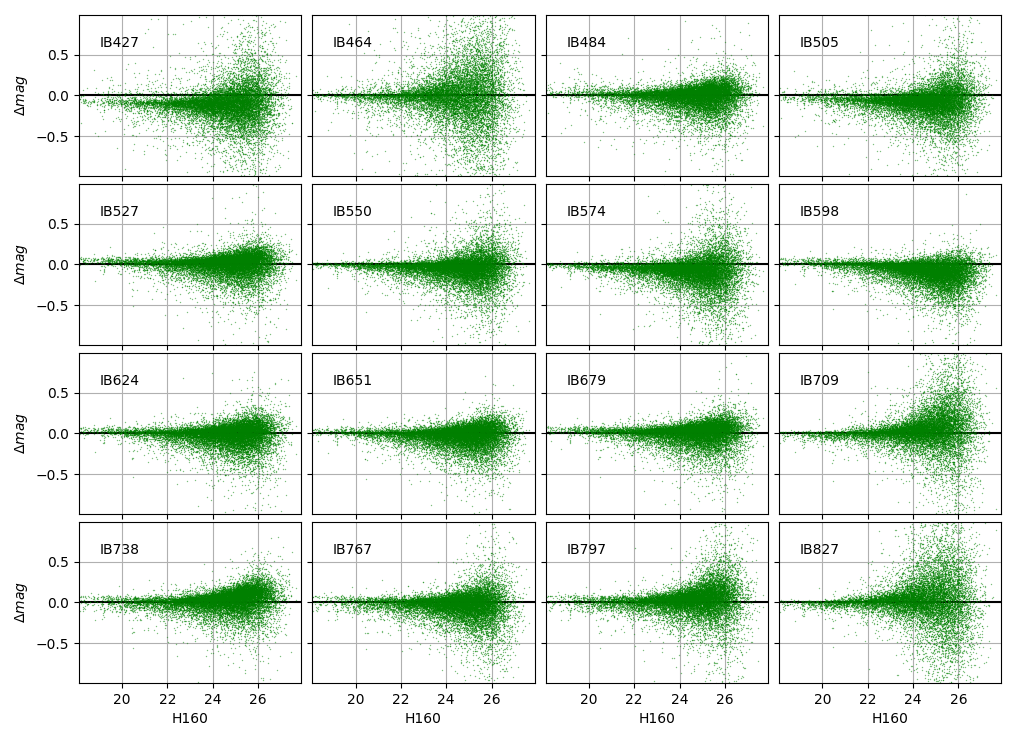}
\caption{Comparison of measured magnitudes between this work and the MUSYC catalog, for 16 reference medium bands.}
\label{comp_musyc}
\end{figure}


\newpage
\section{Description of the ASTRODEEP-GS43 catalogue} \label{relcat}

The sources in the catalogue are ordered by their IDs. The first 34930 are the $H$-detected ones from G13; the subsequent ones are the new $Ks$ and IRAC-detected sources discussed in Sect. \ref{kdet}.

We release two separate files, in ASCII format. The first one, named \texttt{ASTRODEEP-GS43\_phot.cat}, includes IDs, coordinates (RA and DEC), and the 43 bands photometry (not cleaned for the photo-$z$ estimation described in Sect. \ref{zpc}); the passbands are listed in order of increasing wavelength, and we provide total fluxes and corresponding uncertainties in $\mu$Jy. 
The second one, \texttt{ASTRODEEP-GS43\_phys.cat}, includes:
\begin{itemize}
    \item the best redshift estimation, i.e. the spectroscopic $z$ when available (including the original reference), or the median of the three photometric $z$'s obtained with \textsc{LePhare}, \textsc{EAzY} and \textsc{z-phot} otherwise, along with the three estimates, and their standard deviation;
    \item two physical parameters obtained from the best fitting template SED, namely stellar mass and SFR, with 1$\sigma$ uncertainties (given as lower and upper values of the 68\% confidence interval), plus the quality flag described in Sect. \ref{physpar}, for two SFH models, namely exponentially declining (\texttt{tau}) and delayed exponentially declining (\texttt{deltau}; see Sect. \ref{photoz} for details).
\end{itemize}


\end{document}